\documentclass[twocolumn,aps,prb]{revtex4-2}
\usepackage{graphicx,bm,amsmath,amssymb,soul,xcolor,tikz,pgfplots,booktabs,mathtools,titlesec}

\def\<{\langle}
\def\>{\rangle}
\def\br{{\bf r}}
\def\bk{{\bf k}}
\def\bq{{\bf q}}

\def\bR{{\bf R}}
\def\bG{{\bf G}}
\def\a{{\alpha}}
\def\b{{\beta}}
\def\k{\kappa}
\def\w{\omega}
\def\btau{{\bm \tau}}
\def\ve{\varepsilon}
\def\F{Fr\"ohlich}
\def\d{\delta}
\def\aalpha{\lambda}
\def\bbeta{\mu}
\def\ggamma{\rho}
\def\bqin{{\bf q}_\parallel}

\def\bQ{{\bf Q}}
\def\Qin{Q_\parallel}

\def\bGin{{\bf G}_\parallel}

\def\D{\partial}

\begin{document}

\title{Polarons in two-dimensional atomic crystals}

\author{Weng Hong Sio} 
\affiliation{Institute of Applied Physics and Materials Engineering, University of Macau, 
Macao SAR 999078, P. R. China}
\affiliation{Oden Institute for Computational Engineering and Sciences, The University of 
Texas at Austin, Austin, Texas 78712, USA}

\author{Feliciano Giustino}
\email{fgiustino@oden.utexas.edu}
\affiliation{Oden Institute for Computational Engineering and Sciences, The University of 
Texas at Austin, Austin, Texas 78712, USA}
\affiliation{Department of Physics, The University of Texas at Austin, Austin, Texas 
78712, USA}

\date{\today}

\begin{abstract} 
The polaron is the archetypal example of a quasiparticle emerging from the interaction between fermionic 
and bosonic fields in quantum field theory. In crystalline solids, polarons are formed when electrons and holes 
become dressed by the quanta of lattice vibrations. While experimental signatures of polarons in bulk three-dimensional 
materials abound, only rarely have polarons been observed in two-dimensional atomic crystals. Here, we shed light 
on this asymmetry by developing a quantitative \textit{ab initio} theory of polarons in atomically-thin crystals.
Using this conceptual framework, we unravel the real-space structure of the recently-observed hole polaron in 
hexagonal boron nitride, we discover an unexpected critical condition for the existence of polarons in two-dimensional 
crystals, and we establish the key materials descriptors and the universal laws that underpin polaron physics 
in two dimensions.
\end{abstract}

\maketitle

Polarons in solids are quasiparticle excitations resulting from the interaction between electrons and 
phonons \cite{Franchini2021}. Intuitively, polarons can be understood as composite particles where electrons are 
accompanied by a surrounding distortion of the crystal lattice. In the presence of weak electron-phonon interactions, 
polarons behave like conventional Bloch waves, only with heavier effective masses. In the presence of strong 
interactions, on the other hand, polarons become narrow wavepackets and their spatial localization profoundly
alters the transport, electrical, and optical  properties of the host material \cite{Emin2012, Alexandrov2012}.

During the last few years, a number of ground-breaking experimental observations of polarons have been made,
including in quantum materials \cite{Riley2018,Wang2016,Chen2015,Cancellieri2016}, water-splitting photocatalysts 
\cite{Pastor2019,Moser2013,Baldini2020}, photovoltaic perovskites \cite{Miyata2018, Guzelturk2021,Miyata2017b} 
and related double perovskites \cite{Wu2021}. The common denominator to these studies is that they focus on 
three-dimensional (3D) bulk materials. 

In stark contrast with these observations, direct experimental evidence of polarons 
in strictly two-dimensional (2D) materials is scarce. 
Also, from a theoretical standpoint, little is known about polarons in 2D atomic crystals. It is currently 
unknown whether polarons can form in 2D materials, whether they are localized and to what extent, and 
how do they respond to external probes. 
This asymmetry is puzzling when one considers the tremendous 
progress that the field of 2D materials has seen over the past decade and a half \cite{Novoselov2005, Novoselov2016}. 
Two notable exceptions are recent studies of few-layer hexagonal boron nitride (h-BN) on graphene \cite{Chen2018}, 
and monolayer molybdenum disulfide (MoS$_2$) \cite{Kang2018}, where mass enhancement and phonon satellites in 
angle-resolved photoelectron spectra (ARPES) have been observed. It has been proposed that these effects 
arise from the formation of \F\ hole polarons at the h-BN/graphene interfaces, and Holstein polarons in MoS$_2$. 
A definitive assignment of the nature of polarons in these systems would require one to probe the real-space structure 
of the polaron wavefunction or the strain field~\cite{Guzelturk2021}, but this information is currently 
inaccessible via ARPES which is a momentum-space probe.

Here, we shed light on the nature and existence of polarons in atomically-thin 2D crystals by asking how the 
energetics and localization of polaron quasiparticles evolves from 3D to 2D, what are the key materials parameters
that drive polaron formation in 2D, and why 2D polarons are apparently more difficult to detect than in bulk materials. 
To answer these questions, we proceed in two steps: first, we focus on h-BN as a case study, and we compute and 
compare polarons in the bulk crystal and the monolayer. Second, we generalize our findings by developing an 
exactly-solvable \textit{ab initio}-based model of polarons in 2D materials. This step allows us to examine a 
broader class of compounds, and to discover hitherto-unknown laws in the physics of polarons in two dimensions.

\raggedbottom

\begin{figure*}
  \centering
  \includegraphics[width=\textwidth]{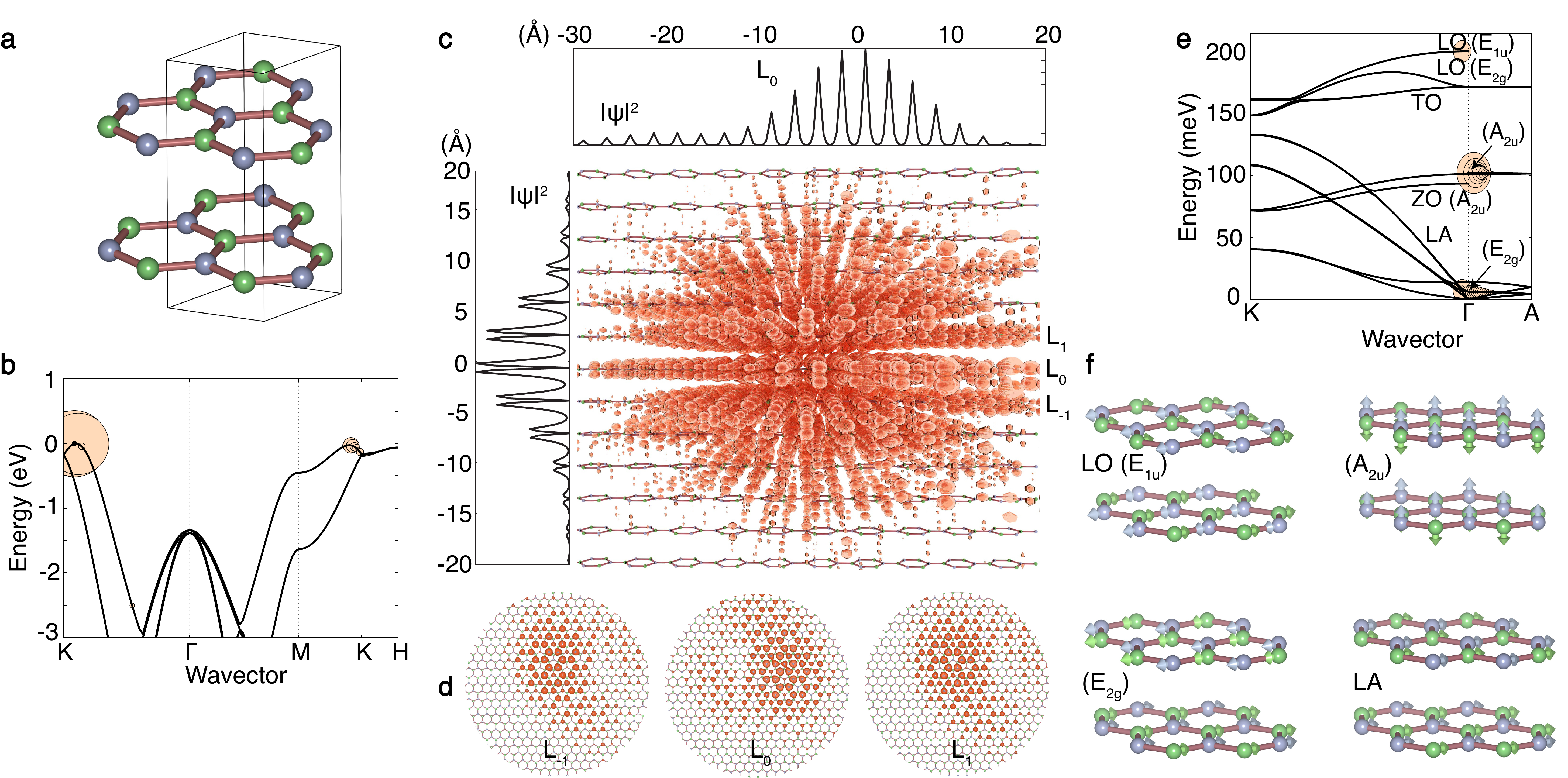}
  \caption{\label{fig1}
   Large hole polaron in bulk h-BN. 
   (a) Ball-stick model of bulk h-BN with B and N in green and blue, respectively. (b) Valence 
   band structure of bulk h-BN. The Fourier amplitudes of the hole polaron wavefunction are superimposed to 
   the bands as circles with radii proportional to $|A_{n\bk}|^2$. (c)  Isosurface plot of the hole 
   polaron density $|\psi(\br)|^2$ for bulk h-BN for a 24$\times $24$\times$8 supercell (9216 atoms). 
   An enlarged view of this wavefunction is shown in Supplemental Fig.~S2. The 
   left and top panels represent one-dimensional profiles of this density, obtained as the planar average 
   along the $c$ axis (left), and as a slice along a line passing through the center of the polaron (top), 
   respectively. (d) Layer-resolved polaron wavefunction, showing the alternating concentration of 
   charge density from one layer to the next. (e) Phonon dispersion relations of bulk h-BN. The 
   Fourier amplitudes of the polaronic distortion are superimposed to the bands as circles with radii
   proportional to $|B_{\bq\nu}|^2$. (f) Atomic displacement patterns of the $E_{1u}$, $A_{2u}$, 
   $E_{2g}$, and LA modes that contribute the most to the polaronic distortion in bulk h-BN.}
\end{figure*}

Our present analysis relies on two recent advances: (i) an \textit{ab initio} computational method for determining 
polaron formation energies, wavefunctions, and atomic displacements without explicit supercell calculations 
\cite{Sio2019a,Sio2019b}; (ii) a unified first-principles formulation of long-range polar electron-phonon couplings
in 3D and 2D \cite{Sio2022}.  We express the polaron wavefunction as $\psi(\br) = N^{-1/2}\sum_{n\bk} A_{n\bk} 
u_{n\bk}(\br)e^{i\bk\cdot\br}$, where $\br$ is the position, $u_{n\bk}$ is the Bloch-periodic component of a 
single-particle electron wavefunction for the band $n$ and wavevector $\bk$, and the summation runs over a 
uniform grid of $N$ points in the Brillouin zone. The coefficients $A_{n\bk}$ describe the contribution of the 
Bloch state $\psi_{n\bk}$ to the polaron. Similarly, we express the displacements of the atom $\kappa$ in the 
unit cell defined by the lattice vector $\bR$ as $\Delta \btau_{\kappa}(\bR) = -2N^{-1}\sum_{\bq \nu} B^*_{\bq\nu} 
(\hbar/2M_\kappa \omega_{\bq \nu})^{\frac{1}{2}} {\bf e}_{\kappa, \bq \nu} e^{i \bq \cdot \bR}$, where $M_\kappa$ 
is the atomic mass, $\bf{e}_{\kappa, \bq \nu}$ is the normal vibrational mode for branch $\nu$ and wavevector $\bq$, 
and $\omega_{\bq\nu}$ is the corresponding vibrational frequency. The coefficients $B_{\bq \nu}$ provide the 
contributions of each normal mode to the polaronic lattice distortion. The vectors $A_{n\bk}$ and $B_{\bq\nu}$ 
are obtained by solving the \textit{ab initio} polaron equations \cite{Sio2019b,Sio2019a}:
\begin{eqnarray} 
 \label{eq.2} &&\frac{2}{N}\sum_{\bq m\nu}B_{\bq\nu}\,g^*_{mn\nu}(\bk,\bq)\,A_{m\bk+\bq}=(\ve_{n\bk}-\ve)\,A_{n\bk},\\
 \label{eq.3} &&B_{\bq\nu}=\frac{1}{N}\sum_{mn\bk}A^*_{m\bk+\bq}\,\frac{g_{mn\nu}(\bk,\bq)}{\hbar\w_{\bq\nu}}\,A_{n\bk}.
\end{eqnarray}
In these expressions, $\ve_{n\bk}$ denotes a single-particle electron eigenvalue, $g_{mn\nu}(\bk,\bq)$ is the 
electron-phonon matrix element connecting the state $\psi_{n\bk}$ with the state $\psi_{m\bk+\bq}$ via the phonon 
of frequency $\w_{\bq\nu}$ \cite{Giustino2017}, and $\ve$ is the polaron eigenvalue. Equations~\eqref{eq.2} and 
\eqref{eq.3} are solved iteratively, and polaron energies are extrapolated to the limit of infinitely-large supercell 
by densifying the Brillouin zone sampling, as shown in Supplemental Figs.~S1-S4. Importantly, this method
does not suffer from the self-interaction error of density functional theory, as discussed in Supplemental Note~1. 
All calculations are performed 
using Quantum ESPRESSO \citep{Giannozzi2017}, Wannier90 \citep{Mostofi2014}, and EPW \citep{Ponce2016}. A detailed 
description of this approach can be found in Ref.~\citep{Sio2019b}, and the computational setup including a new
formulation of the electron-phonon matrix element in 2D is outlined in the Computational Methods section
of the Supplemental Material.

\begin{figure*}
  \centering
  \includegraphics[width=0.9\textwidth]{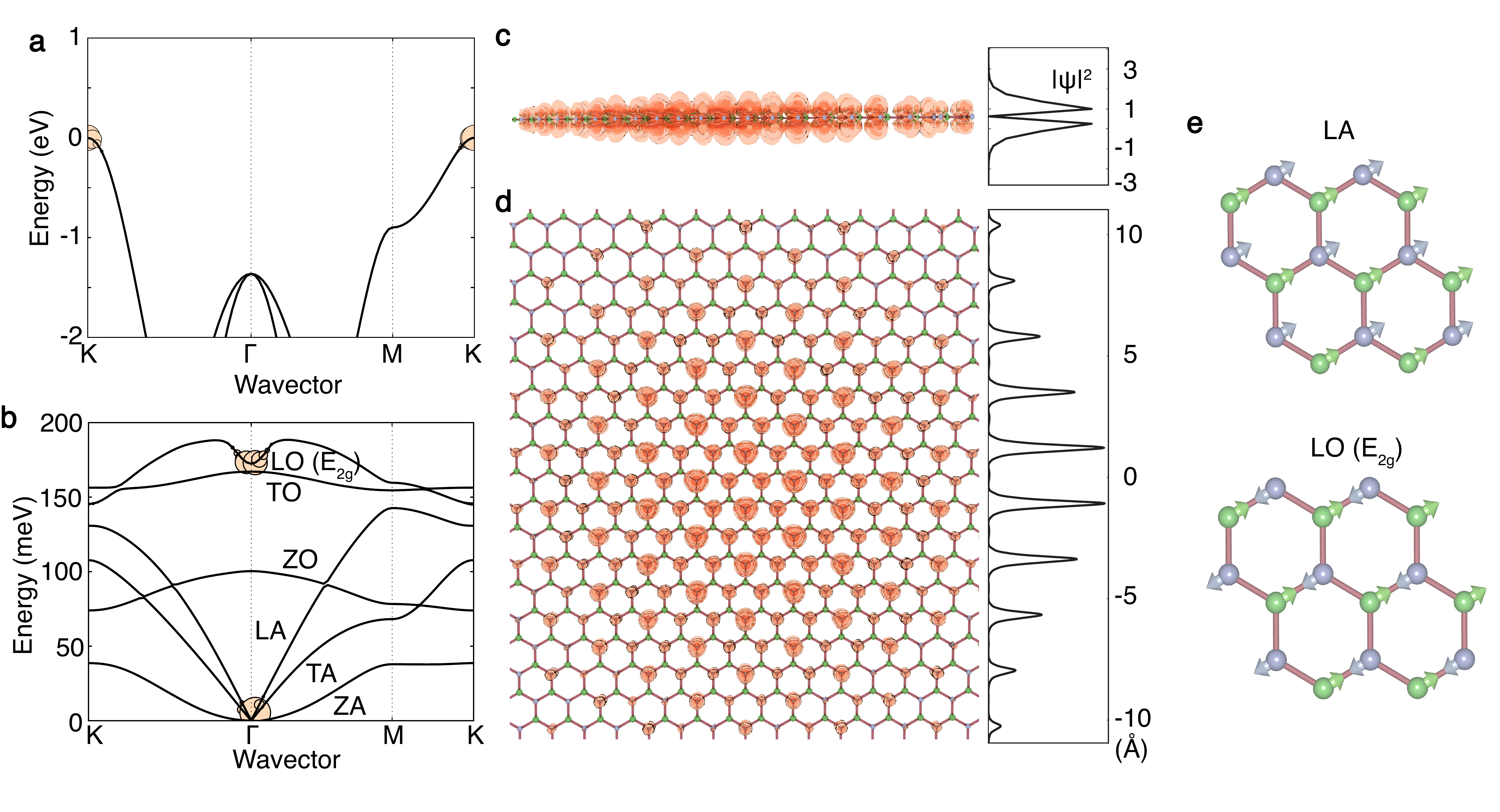}
  \vspace{-15pt}
  \caption{
  \label{fig2}
   \F\ polaron in monolayer h-BN. 
   (a) Valence band structure of monolayer h-BN. The Fourier amplitudes of the hole polaron wavefunction 
   are superimposed to the bands as circles with radii proportional to $|A_{n\bk}|^2$. (b) Phonon dispersion 
   relations of monolayer h-BN. The Fourier amplitudes of the polaronic distortion are superimposed to the bands as 
   circles with radii proportional to $|B_{\bq\nu}|^2$. 
   Note that the LO-TO splitting is almost entirely suppressed, as expected from the reduced dimensionality 
   \cite{Sohier2017}. We see a small residual splitting because the calculations are performed using a large but
   finite supercell in the $c$-axis direction, cf. Computational Methods section of the Supplemental Material.
   (c)  Top view of the hole polaron density $|\psi(\br)|^2$ 
   for monolayer h-BN for a 26$\times$26$\times$1 supercell (676 atoms), and one-dimensional profile obtained as the 
   planar average along the axis perpendicular to the monolayer. (d) Side view of the hole polaron density, and 
   one-dimensional profile obtained as a slice along a line passing through the center of the polaron. An enlarged 
   view of this wavefunction is shown in Supplemental Fig.~S4. (e) Atomic displacement patters of the 
   LO E$_{2g}$ and LA modes that provide the dominant contribution to the polaronic distortion in monolayer h-BN.
}
\end{figure*}

\raggedbottom

In Fig.~\ref{fig1} we illustrate our results for the hole polaron in bulk h-BN. Figure~\ref{fig1}(a) shows
the structure of bulk h-BN in the AA$^\prime$ stacking. In this configuration, the B atoms in one layer lie 
directly above the N atoms in the layer underneath. Figure~\ref{fig1}(b) shows the valence bands of bulk h-BN. 
The top of the bands is near the $K$ point and derives from N-$2p_z$ orbitals. Upon removing one electron 
and letting the atoms adjust around the hole, a polaron forms as shown in Fig.~\ref{fig1}(c). The polaron 
wavefunction extends over 10 unit cells in the $ab$ plane, and spans 7 atomic layers along the $c$ direction, 
therefore this is a large hole polaron. 
Incidentally, we note that performing this calculation without employing the \textit{ab initio} polaron
equations Eqs.~(1)-(2) would have required supercells of at least 36,000 atoms [Supplemental Fig.~S1(a)], 
which is currently beyond reach using hybrid functionals.
Despite the highly anisotropic nature of h-BN, the wavefunction 
is found to be nearly spherical in shape, with full-width-at-half-maximum of 13~\AA\ along the $c$ axis 
and 15~\AA\ in the $ab$ plane. This can be seen in the one-dimensional charge density profiles in 
Fig.~\ref{fig1}(c). From the layer-resolved plots in Fig.~\ref{fig1}(d), we can also see that the hole 
density tends to accumulate on opposite sides from one layer to the next. This intriguing pattern 
originates from the fact that the transverse-optical (TO) $E_{2g}$ interlayer shear mode [Fig.~\ref{fig1}(f)]
generates macroscopic dipoles with signs alternating between layers, thereby pulling the 
hole in a zig-zag pattern. The formation energy of this polaron, as measured from the fully-delocalized 
ground state, is $E = -$13.6~meV, therefore we have a weak polaronic renormalization of the valence band 
energy. 

Figures~\ref{fig1}(b) and (e) show the decomposition of the polaron wavefunction and the accompanying 
atomic displacements into the underlying Bloch states, as obtained by overlaying the coefficients $|A_{n\bk}|^2$
and $|B_{\bq\nu}|^2$ on the electron and phonon bands. The hole wavefunction derives from the ring 
of valence band maxima enclosing the $K$ point \citep{Wickramaratne2018}. The distribution in reciprocal 
space is very narrow, consistently with the large size of the wavefunction in real space.  Analogous trends 
have been observed for excitons in h-BN \cite{Arnaud2006,Wirtz2008,Galvani2016,Zhang2022}. Only long-wavelength 
phonons are involved in the formation of the polaron, in particular a longitudinal optical (LO) phonon, two 
TO phonons, and one longitudinal acoustic (LA) phonon. In Fig.~\ref{fig1}(f) we see that the LO phonon 
is an $E_{1u}$ bond-stretching mode propagating in-plane at 201~meV (38\%); the TO phonons are the the 
$A_{2u}$ $c$-axis sliding mode of the B and N sublattices at 102~meV (27\%) and the $E_{2g}$ interlayer 
shear mode at 6~meV (4\%), both of which propagate out-of-plane; the LA phonon (31\%) is an in-plane 
compression wave \cite{Gil2020}. Since the contributions of LO, TO, and LA phonons are comparable in size, 
%we can say that bulk h-BN hosts hybrid \F-Holstein-acoustic polarons.
the appropriate classification of polarons in bulk h-BN is as hybrid \F-Holstein-acoustic polarons.
The numerical values provided above will no doubt improve when many-body calculations of 
electron-phonon matrix elements \cite{Giustino2017} will become widely available,
such as for example the recent GWPT method \cite{Li2019prl}. However, we expect 
such improvements to be of the order of 5\% or less based on the error analysis reported in Supplemental 
Note~S2.

Next we discuss polarons in monolayer h-BN. The 2D version of h-BN has received considerable attention 
as a versatile platform for hyperbolic phonon polaritons \cite{Dai2014}, single-photon emitters
\cite{Tran2016}, and deep UV emission \cite{Cassabois2016,Elias2019}, but the possible role of polarons 
in these applications is currently unknown. Figure~\ref{fig2}(a) shows the band structure of the monolayer,
with the top of the valence bands located at the $K$ point of the hexagonal Brillouin zone. Similarly to
the case of bulk h-BN, the polaron wavefunctions primarily draws weight from the $K$ valleys.  Isosurface 
plots of the polaron wavefunction are shown in Fig.~\ref{fig2}(c) and (d), with views from the side and from 
the top, respectively, as well as the corresponding one-dimensional profiles. The atomic displacements of 
this polaron are shown in Supplemental Fig.~S3. The wavefunction consists of N-$2p_z$ orbitals modulated 
by a Gaussian-like envelope, and extends over 9 unit cells with a full-width at half maximum of 10~\AA. 
This wavefunction is more localized than in the case of bulk h-BN, and accordingly the formation energy is 
higher than in bulk, $E = -$15.9~meV. The structure of the polaron wavefunction is reminiscent of the 
excitonic wavefunction in monolayer h-BN obtained via the Bethe-Salpeter approach~\cite{Galvani2016}. 
This similarity reflects the common root of both polarons and excitons in the Coulomb interaction between 
charged excitations.

Unlike in bulk h-BN, in monolayer h-BN there is no contribution from the interlayer shear mode
and from the $c$-axis sublattice sliding mode, which only exist when more than one layer is present.  
This is shown in Fig.~\ref{fig2}(b). The only vibrational modes contributing to polaron formation are the 
long-wavelength $E_{2g}$ bond-stretching LO phonon 
at 172~meV (76\%), and the LA phonon (24\%). The atomic displacement patterns of these modes are 
shown in Fig.~\ref{fig2}(e). Based on these results, we infer that the polaron in monolayer h-BN is dominated 
by polar phonon coupling, and therefore it can be classified as a large \F\ polaron. 

\raggedbottom

Our present results indicate that the \F\ interaction plays an important role both in bulk and
in monolayer h-BN. This finding is in line with the recent observation of polaron fingerprints at the 
interface between few-layer h-BN and graphene \cite{Chen2018}. In Ref.~\onlinecite{Chen2018} ARPES
satellites were observed $\sim$210~meV below the valence quasiparticle peak. Our calculations support the 
assignment of these features to polarons with significant \F\ character, on the grounds that the LO mode 
at $\sim$200~meV in bulk h-BN [Fig.~\ref{fig1}(e)] and its counterpart in monolayer h-BN [Fig.~\ref{fig2}(b)] 
contribute significantly to the formation of the polaron.

If we compare the above results for bulk and monolayer h-BN, we find that the polaron is more stable in the
monolayer. This trend is consistent with the observation of stronger exciton binding in van der Waals materials
in the monolayer limit \cite{Ugeda2014,Olsen2016}. Closer inspection, however, indicates that the analogy 
between polarons and excitons does not go further: while the exciton binding energy increases by almost an 
order of magnitude, from 130~meV in bulk h-BN to 0.7$\pm$ 0.2~eV in the monolayer \cite{Cassabois2016,Paleari2018,
Roman2021}, the increase in polaron formation energy is of only 17\% (from 13.6~meV to 15.9~meV). This 
surprising result calls for a more in-depth analysis. In the following, we investigate how dimensionality 
determines the nature of polarons by deriving an exactly-solvable model.

\raggedbottom

\begin{figure*}
  \centering
  \includegraphics[clip,trim=1.5cm 14.9cm 1.6cm 2.0cm,width=0.9\textwidth]{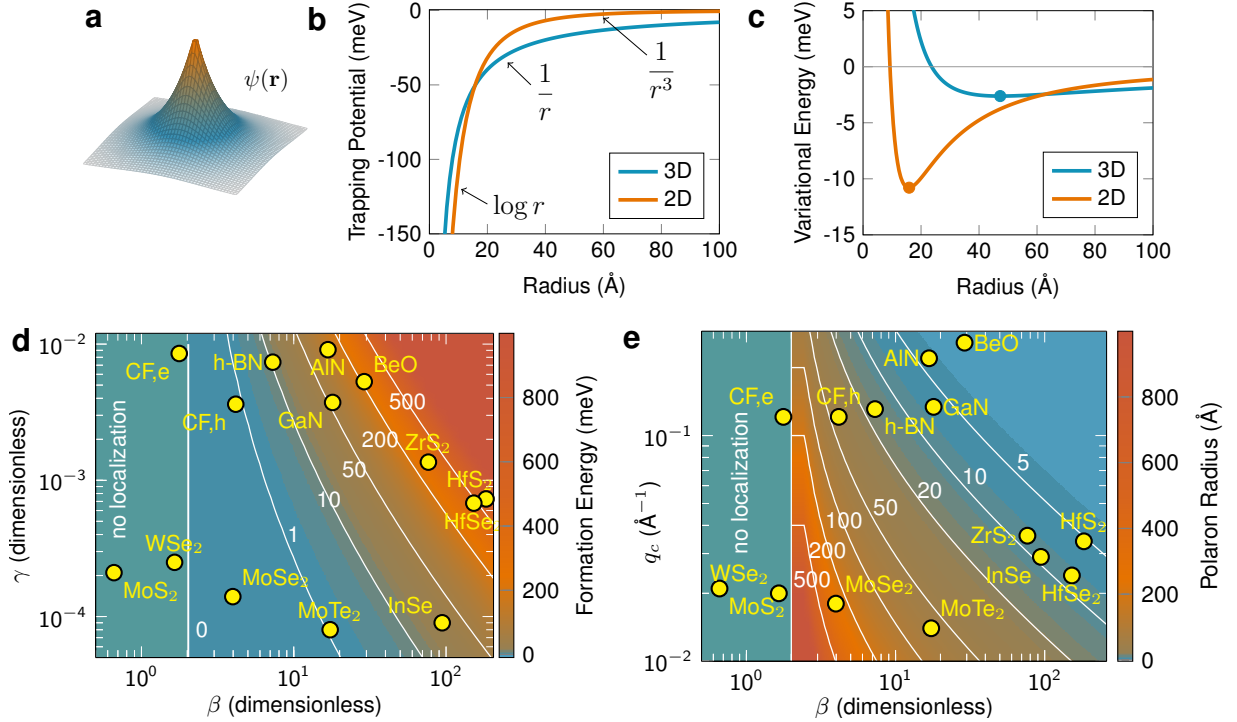}
  \caption{\label{fig3}
   Continuum model of polarons in two-dimensional atomic crystals.
   (a) Schematic illustration of the trial variational wavefunction for the 2D \F\ polaron model of 
    Eq.~\eqref{eq.E}.  (b) Effective polaron self-trapping potentials in three dimensions (cyan) and 
    in two dimensions (orange).  For definiteness we use the in-plane dielectric constants of bulk h-BN 
    ($\epsilon_0 =$~5.0 and $\epsilon_\infty =$~6.9), hole effective mass $m^* =$~0.65~$m_{\rm e}$ of monolayer h-BN,  
    and set the thickness of the monolayer to the bulk interlayer distance, $d =$~3.12~\AA. (c) Energy of polarons 
    in 3D (cyan) and 2D (orange) as a function of the polaron radius $r_p$. The 3D curve is from the Landau-Pekar 
    model, the 2D curve is from Eq.~\eqref{eq.ef.var}. The dots at the minimum of each curve indicate stable 
    polaron solutions. We employ the same parameters for h-BN as in panel (b). (d) 
    Formation energy map of hole polarons in 2D crystals, 
    as a function of the materials descriptors $\b = \epsilon_{\rm ion} m^* d/m_{\rm e} a_0$, 
    $\gamma = q_c^2 a_0^2 m/m^*$, and $q_c d = 4 \epsilon_{\infty} /(2\epsilon_{\infty}^2 \!-\!1)$. Polarons do 
    not form when $\beta \le 2$ (green area). Isolines indicate energies in meV.  The yellow disks correspond to hole
    polarons for the 2D materials reported in Supplemental Table S3. In the case of fluorographene, we report
    both hole polaron (CF,h) and electron polaron (CF,e). (e) Map of hole polaron radii 
    in 2D crystals, as a function of the materials descriptors $\b$ and $q_c$. The size of the polaron becomes 
    infinite at $\beta =2$. Isolines indicate radii in \AA. Disks represent the same materials reported in (d).}
\end{figure*}

For simplicity we focus on the \F\ interaction, which dominates the polaron in monolayer h-BN, and we
neglect the acoustic phonon contribution. Therefore the following results represent lower bounds to 
the formation energies.
Several studies of \F\ polarons in 2D have been reported, shedding
light on important aspects of the many-body physics of polarons \cite{Franchini2018,Peeters1985,Ercelebi1987,
Jalabert1989,Titantah2001}. However, in all this prior work the authors considered 2D sheets without thickness,
and this choice leads to unrealistically strong \F\ interactions that diverge at long wavelength \cite{Jalabert1989}. 
Conversely, multiple \textit{ab initio} studies showed that the coupling matrix elements of real 2D crystals are finite 
at long wavelength \cite{Kaasbjerg2012,Sohier2016,Deng2021,Sio2022}. Starting from this observation, we use
the following expression for the \F\ matrix element, which we recently derived from first principles \cite{Sio2022}:
 \begin{equation}\label{eq.g}
 g(q) = i\left[ E_{\rm Ha}\frac{\pi}{2} \frac{a_0 d}{A} \, \hbar\w\, \epsilon_{\rm ion} \right]^{\!1/2}\!\!\!
 \frac{q_c}{q_c+q}.
 \end{equation} 
This matrix element describes the \F\ interaction in a homogeneous 2D layer of thickness $d$ and unit cell 
area $A$, with a single dispersionless LO phonon of frequency $\omega$. The dielectric screening is uniform 
and isotropic within this layer, and $\epsilon_{\rm ion} = \epsilon_0-\epsilon_\infty$ is the difference between 
the static relative permittivity $\epsilon_0$ and the high-frequency relative permittivity $\epsilon_\infty$, 
i.e. the ionic contribution to the dielectric constant. $E_{\rm Ha}$ is the Hartree energy, $a_0$ the Bohr radius, 
and $q_c$ is a characteristic wavevector defined by $q_c d = 4 \epsilon_{\infty}/(2\epsilon_{\infty}^2 \!-\!1)$ 
\cite{Sio2022}. The key difference between Eq.~\eqref{eq.g} and the standard \F\ matrix element in 3D is the presence 
of $q_c$ at the denominator, which eliminates the characteristic long-wavelength singularity that one encounters in 
bulk crystals.

Using Eq.~\eqref{eq.g} inside Eqs.~\eqref{eq.2}-\eqref{eq.3}, and considering a single parabolic band with
effective mass $m^*$, we transform the equations into real space to find the polaron formation energy:
 \begin{equation}\label{eq.E}
 E = \!\frac{\hbar^2}{2m^*}\!\!\!\int \!d\br \,|\nabla \psi|^2 +
 \!\frac{1}{2}\!\int \!\!d\br \,d\br'\, |\psi(\br)|^2 V_{\rm 2D}(\br-\br') |\psi(\br')|^2\!,
 \end{equation}
where $\psi$ is the polaron wavefunction and $\br$ is the 2D coordinate. The effective potential $V_{\rm 2D}$ is
given by:
 \begin{equation}\label{eq.V2d.1}
 V_{\rm 2D}(r) = - \frac{1}{2}E_{\rm Ha}\, a_0\, d q_c^2\, \epsilon_{\rm ion} \,\phi(q_c r),
 \end{equation}
where $r=|\br|$ and the dimensionless function $\phi$ is:
 \begin{equation}\label{eq.V2d.2}
 \phi(s) = s + \frac{\pi}{2}\left[ H_0(s)-Y_0(s)\right] - \frac{\pi}{2} s\left[  H_1(s)- Y_1(s)\right].
 \end{equation}
In this equation, $H_0$ and $H_1$ are Struve functions, $Y_0$ and $Y_1$ are Bessel functions of the second kind.
The derivations of Eq.~\eqref{eq.E} and Eqs.~\eqref{eq.V2d.1}-\eqref{eq.V2d.2} are provided in Supplemental Notes~3
and 4, respectively. The first and second term on the right-hand side of Eq.~\eqref{eq.E} represent the kinetic 
energy of the polaron, which favors delocalization; and the potential energy of phonon-mediated self-trapping, 
which favors localization. Equation~\eqref{eq.E} can be considered as the generalization of the Landau-Pekar 
model \cite{Landau1933,Pekar1946,Devreese2020} to 2D systems with realistic \F\ interactions.
In order to highlight similarities and differences between the current model for 2D polarons and the
Landau-Pekar model, in Supplemental Table~S2 we compare side-by-side the key equations that define these
two models.

Figure~\ref{fig3}(b) shows the trapping potential $V_{\rm 2D}$. To make contact with our \textit{ab initio}
calculations for h-BN, we use the dielectric constants and effective mass of bulk and monolayer h-BN, respectively,
and we set the monolayer thickness to the interlayer distance in bulk h-BN. For comparison, in the same figure we plot 
the corresponding trapping potential in 3D, which has the standard Coulomb form $V_{\rm 3D}(r) = - E_{\rm Ha} a_0 
(\epsilon_\infty^{-1} -\epsilon_0^{-1})/r$ \cite{Sio2019b,Devreese2009}. Unlike in the bulk case, the potential well 
in 2D is deeper and short-ranged: At short distance ($r\ll q_c^{-1}$) the potential diverges as $\log r$, and at 
long distance ($r \gg q_c^{-1}$) it behaves as $r^{-3}$ \cite{Spanier1987}. Equation~\eqref{eq.V2d.2} is 
reminiscent of the the Rytova-Keldysh potential \cite{Rytova1967,Keldysh1979} used in the study of excitons 
in 2D materials \cite{Cudazzo2011}, except that it contains additional terms and it is short-ranged.

\raggedbottom

To analyze the consequences of the short-ranged nature of $V_{\rm 2D}$, we perform a variational calculation of 
the polaron formation energy. After replacing the hydrogenic ansatz $\psi(\br) = (2/\pi r_p^2)^{1/2}\exp(-|\br|/r_p)$ 
\cite{Sio2019b,Pekar1946,Devreese2009} [cf. Fig.~\ref{fig3}(a)] inside Eq.~\eqref{eq.E}, in Supplemental Note~5
we find:
 \begin{equation}\label{eq.ef.var}
 E = \frac{\hbar^2}{2 m^* r_p^2} - \frac{\epsilon_{\rm ion}}{4}\frac{e^2 d}{4\pi \varepsilon_0  r_p^2}\, f(q_c r_p).
 \end{equation}
In this expression, the polaron radius $r_p$ serves as the variational parameter, and the dimensionless 
function $f$ is given in footnote \footnote{The dimensionless function $f$ appearing in Eq.~\eqref{eq.ef.var} 
is $ f(x) = x^2[ x^6 -\pi x^5 +28x^4 -40 \pi x^3 - 16 (1+20\log 2) x^2 +240\pi x + 64(4 \log 2 -7) +64 
(5x^2-4)\log x ]/ (x^2+4)^4$}. The two terms on the right-hand side of Eq.~\eqref{eq.ef.var} correspond to 
the kinetic and potential energies in Eq.~\eqref{eq.E}, respectively. The competition between these terms 
can lead to localized polarons being more stable than delocalized Bloch electrons. 
Figure~\ref{fig3}(c) shows how the energy $E$ depends on the 
variational parameter $r_p$ for the case of monolayer h-BN. For comparison, we also show the formation 
energy for bulk h-BN, obtained from the Landau-Pekar model \cite{Sio2019b,Devreese2009}. In agreement 
with our \textit{ab initio} calculations, the 2D polaron is found to be more stable and more localized 
than its 3D counterpart. Furthermore, the formation energies obtained from these models, $E = -$10.8~meV 
for monolayer h-BN and $-$2.6~meV for bulk h-BN, are in good agreement with our first-principles results. 
Indeed, since in our \textit{ab initio} calculations the LO modes account for 76\% and 38\% of the formation 
energies in monolayer and bulk h-BN, respectively, including only the \F\ interaction in these calculations 
would yield formation energies of $-$12.1~meV and $-$5.2~meV, very close to our analytical model.

The modest increase in formation energy from 3D to 2D in h-BN can now be rationalized 
starting from an analogy with excitons. In the case of excitons, the Coulomb interaction in 2D is stronger than
in 3D due to the lack of bulk screening~\cite{Cudazzo2011}. 
Extrapolating this reasoning to polarons would lead to the conclusion that
formation energies should also increase significantly from 3D to 2D. However, inspection
of Eqs.~\eqref{eq.g} and \eqref{eq.ef.var}
reveals that polaron formation is driven by the ionic contribution to the dielectric screening
($\epsilon_{\rm ion}=\epsilon_0-\epsilon_\infty$), which is significantly reduced in 2D. 
To see this, let us imagine a bulk system where all layers polarize to trap the
electron; if we remove all layers except one, the Coulomb interaction strength between the
electron and the surrounding atoms increases due to the reduced screening (as for excitons), however the
trapping potential weakens due to the removal of polarizable layers (unlike excitons). These two effects partially compensate,
leading to a milder increase of formation energy from 3D to 2D as compared to excitons.

Besides providing us with a compact conceptual framework to interpret detailed \textit{ab initio} calculations, 
the model of Eq.~\eqref{eq.ef.var} offers fundamental new insight into the difference between polarons in 2D 
and 3D. In 3D the self-trapping potential energy goes as $r_p^{-1}$ \cite{Sio2019b}, therefore the interplay 
between the repulsive kinetic term (which goes as $r_p^{-2}$) and the attractive Coulomb term always leads 
to a minimum in the energy landscape. By consequence, in 3D polarons will form no matter how weak the electron-phonon coupling 
\cite{Feynman1955,Prokofev1998, Hahn2018,Sio2019b,Devreese2020}. The situation is fundamentally different in 
2D. In fact, at large polaron radii, the function $f$ in Eq.~\eqref{eq.ef.var} becomes a constant [cf. 
Supplemental Fig.~S5], therefore 
both the attractive and the repulsive terms scale as $r_p^{-2}$.  In this limit, polarons can form only if the 
prefactor of the attractive term exceeds that of the repulsive term. By requiring this inequality to hold, 
we obtain a critical condition for the existence of polarons in 2D:
  \begin{equation}\label{eq.crit}
 %    \epsilon_{\rm ion} \frac{m^*}{m_{\rm e}} \frac{d}{a_0} > 2 ,
     \epsilon_{\rm ion} m^* d > 2\, m_{\rm e}a_0 ,
  \end{equation}
where $m_{\rm e}$ is the bare electron mass. One can verify that this condition must also hold for any radius $r_p$, 
because the function $f$ in Eq.~\eqref{eq.ef.var} is monotonic and bound by $f=1$ from above, as shown in 
Supplemental Fig.~S5. Direct calculations of the formation energy for a broad range of materials parameters, 
as shown in Fig.~\ref{fig3}(d), confirm that 2D polarons do not form when Eq.~\eqref{eq.crit} is not satisfied.
As a further test of the validity of this critical condition, we perform explicit \textit{ab initio}
calculations of fluorographene, using both the \textit{ab initio} polaron equations Eqs.~(1)-(2) and the
self-interaction-corrected density-functional approach of Ref.~\onlinecite{Sio2019b}. In both cases
we do not find stable electron polarons, in agreement with the fact that electrons in fluorographene do not
fulfil the critical condition given by Eq.~\eqref{eq.crit}. This is seen in Fig.~\ref{fig3}(d) and discussed
in more detail in Supplemental Note~7 and Supplemental Fig.~S3.

\raggedbottom

The identification of a critical condition for the existence of \F\ polarons in 2D marks a significant departure 
from earlier work. This condition was previously missed because the \F\ coupling matrix element employed 
in prior work \cite{Franchini2018,Peeters1985,Ercelebi1987,Jalabert1989,Titantah2001}, which describes idealized 
2D sheets without thickness, is unrealistically strong and leads to stable polaronic states at all couplings. 
The use of realistic coupling matrix elements as given by Eq.~\eqref{eq.g} \cite{Sio2022,Sohier2016,
Kaasbjerg2012} reveals a more complex picture, where the laws and descriptors previously identified for 3D materials
\cite{Franchini2021} no longer apply. The existence of a critical condition for polaron
formation in 2D might explain why there have been far fewer experimental observations of polarons in 2D 
crystals than in 3D bulk systems. 

Figures~\ref{fig3}(d) and (e) show how polaron energetics and size vary across the landscape of materials 
parameters $m^*$, $d$, $\epsilon_\infty$, and $\epsilon_{\rm ion}$. The four-dimensional parameters space
can be represented via two-dimensional maps by noticing that the formation energy depends only on the 
dimensionless products $\b = \epsilon_{\rm ion} m^* d/m_{\rm e} a_0$ and $\gamma = a_0^2 m_{\rm e} q_c^2/m^*$, while the 
radius depends only on $\beta$ and $q_c$, as discussed in Supplemental Note~6. 

Explicit expressions for the 
formation energy and size of the polaron can be obtained in the weak-coupling limit.
In Supplemental Note~6 we show that, in this limit, the energy scales as $\gamma (\beta -2)^3$, and the polaron radius 
scales as $[q_c(\b-2)]^{-1}$. These scaling laws align with the expectation that strong ionic screening and 
heavy effective masses, which yield large $\b$ values, lead to more stable and more localized polarons. 
Conversely, when this parameter approaches the critical value $\beta = 2$, the polaron radius diverges and 
fully-delocalized
Bloch states are recovered. The predominant dependence of polaron energy and radius on $\beta$ (as compared to
$\gamma$ and $q_c$) suggest that this descriptor should be considered as the 2D counterpart of the \F\ coupling 
constant $\a$ that is commonly used for bulk materials \cite{Devreese2009}. 

The exactly-solvable model in Eq.~\eqref{eq.ef.var} is particularly useful to perform a rapid assessment 
of polaron energetics and size for several 2D materials.  In Figs.~\ref{fig3}(d) and (e) we include datapoints 
for popular 2D crystals, namely h-BN, MoS$_2$, MoSe$_2$, MoTe$_2$, WS$_2$, WSe$_2$, HfS$_2$, HfSe$_2$, ZrS$_2$,
InSe, as well as fluorographene \cite{Nair2010} and the recently-synthesized monolayer h-AlN \cite{Chang2022},
monolayer h-GaN \cite{Balushi2016}, and monolayer h-BeO \cite{Zhang2021}.
For each compound we determine the descriptors $\beta$, $\gamma$, and $q_c$ using the materials
parameters $m^*$, $d$, $\epsilon_\infty$, and $\epsilon_{\rm ion}$ from Refs.~\onlinecite{Sohier2016, Sio2022, 
Wang2021, Lv2016, Li2019, Ferreira2019, Zhang2014, Laturia2018, Pike2018, Li2020}, as reported in Supplemental 
Table~S3. 
For definiteness we focus on hole polarons, and we report the corresponding data for electron polarons 
in Supplemental Table~S3. We immediately see that group-VI transition-metal dichalcogenides (MoS$_2$, MoSe$_2$, 
MoTe$_2$, WS$_2$, WSe$_2$) do not host stable \F\ polarons. In particular, MoSe$_2$ and MoTe$_2$ fulfill the critical 
condition in Eq.~\eqref{eq.crit}, but the polaron wavefunctions extend over 10-50~nm; it is unlikely that 
quantum coherence between electron and phonons be maintained over such distances in real, defective crystals. 
The other group-VI dichalcogenides MoS$_2$, WS$_2$, and WS$_2$ do not satisfy the critical condition in 
Eq.~\eqref{eq.crit} owing to weak ionic screening, therefore in these cases we do not expect carrier localization. 
This finding is in agreement with recent work \cite{Garcia2019} 
showing that the phonon sidebands observed in ARPES spectra of MoS$_2$ \cite{Kang2018} 
arise from the coupling of non-polar zone-boundary phonons to delocalized Bloch electrons.

On the other hand, 
the data for group-IV transition-metal dichalcogenides as well as h-BeO in 
Figs.~\ref{fig3}(d) and (e) paint a very different 
picture.
With polaron sizes in the range of 1-2~nm and formation energies significantly above 100~meV, 
HfS$_2$, HfSe$_2$, ZrS$_2$, and h-BeO appear to be ideal platforms for investigating strong-coupling polaron physics in 
two dimensions, as we confirmed by explicit \textit{ab initio} calculations for the cases of ZrS$_2$ and h-BeO
(see Supplemental Note~7 and Supplemental Figs.~2, 4, 9, 10, 12).
 Group-IV transition-metal dichalcogenide monolayers are
rapidly emerging as a promising alternative to group-VI dichalcogenides 
for field-effect transistors, batteries, thermoelectricity, and catalysis \cite{Yan2018}, therefore experimental 
studies of polarons should be within reach. Furthermore, the recent synthesis of h-BeO monolayers 
by molecular beam epitaxy \cite{Zhang2021} makes this new material an attractive candidate for flatland
electronics owing to its potential for large-scale growth of stable single-crystalline insulators. These emerging
materials platforms
could be used to test time-resolved strain field mapping of polarons via diffuse X-ray scattering, as recently 
demonstrated for bulk materials \cite{Guzelturk2021}, and even to test exotic mechanism of superconductivity 
such as the bipolaronic pairing scenario \cite{Alexandrov1981} in two dimensions.

Beyond these examples, we anticipate that the materials descriptors and the universal laws identified
here will serve as a unifying conceptual framework to understand, probe, and control polarons 
in two-dimensional crystals, and could lead to a step change in our understanding of emergent 
quasiparticles in low-dimensional materials. In particular, the combination of this framework and these
descriptors of polaron behavior with large
databases of 2D materials obtained from computational exfoliation of experimentally known 
compounds \cite{Mounet2018} as well as other 2D databases \cite{Rasmussen2015}
will enable machine learning and AI approaches for deepening our
current understanding of polarons in 2D, and possibly open the doors to data-driven design of 
two-dimensional polaronic materials.

\acknowledgments

This research is primarily supported by the Computational Materials Sciences Program funded by the U.S. Department of Energy, Office of Science, Basic Energy Sciences, under Award No. DE-SC0020129 (software development, theoretical model, \textit{ab initio} calculations). This research used resources of the National Energy Research Scientific Computing Center, a DOE Office of Science User Facility supported by the Office of Science of the U.S.  Department of Energy under Contract No. DE-AC02-05CH11231. The authors also acknowledge the Texas Advanced Computing Center (TACC) at The University of Texas at Austin for providing additional HPC resources, including the Frontera and Lonestar6 systems, that have contributed to the research results reported within this paper. URL: http://www.tacc.utexas.edu. In the final stage, W.H.S. was supported by the Science and Technology Development Fund of Macau SAR (FDCT) under grant No. 0102/2019/A2 (\textit{ab initio} calculations and data analysis). W.H.S also acknowledges the Information and Communication Technology Office (ICTO) at the University of Macau and the LvLiang Cloud Computing Center of China for providing extra HPC resources, including the High Performance Computing Cluster (HPCC) and TianHe-2 systems.

%%%%%%%%%%%%%%%%%%%%%%%% END OF MAIN TEXT %%%%%%%%%%%%%%%%%%%%%%%%%

\clearpage
\newpage

\renewcommand\theequation{S\arabic{equation}}
\renewcommand{\thesection}{\arabic{section}}
\renewcommand{\thetable}{S\arabic{table}}
\renewcommand\thefigure{S\arabic{figure}}
\titleformat{\section}{\large\bfseries}{\thesection.}{5pt}{}
\renewcommand{\baselinestretch}{1.15}
\setlength{\parindent}{0pt}
\setlength{\parskip}{3pt}
\titlespacing{\section}{0pt}{16pt}{5pt}
\def\bibsection{\section*{References}}
\setcounter{equation}{0}
\setcounter{figure}{0}

\onecolumngrid

\begin{center}
\textbf{\large Supplemental Material for:\\[4pt] Polarons in two-dimensional atomic crystals}
\end{center}

\vspace*{5pt}
Contents:\\

Computational Methods\\
Supplemental Notes 1-7\\
Supplemental Tables S1-3\\
Supplemental Figures S1-S12\\
Supplemental References

\leftskip0pt
\section*{Computational Methods}

All \textit{ab initio} calculations are performed using the Quantum ESPRESSO package \citep{Giannozzi2017} 
(electronic structure and lattice vibrational properties), the wannier90 code \citep{Mostofi2014} (maximally-localized
Wannier functions) and the EPW \citep{Ponce2016} code (interpolation of electron-phonon matrix elements and polarons).
We describe h-BN, ZrS$_2$, fluorographene, and BeO using density functional theory (DFT) in the 
local density approximation (LDA: bulk h-BN) and the generalized gradient approximation (PBE: 
monolayer h-BN, ZrS$_2$, fluorographene and monolayer BeO) \cite{Ceperley1980,Zunger1981,Perdew1996},
and we employ ONCV pseudopotentials \citep{Hamann2013,Schlipf2015}. In the case of h-BN, we use
a planewaves kinetic energy cutoff of 125~Ry. In ground-state calculations and in calculations of
phonon frequencies and eigenmodes, we sample the Brillouin zone using a $\Gamma$-centered uniform grid of 
$14\times 14\times 6$ points and $12\times 12\times 2$ points for bulk and monolayer h-BN, respectively. 
All lattice vectors and internal coordinates are optimized before proceeding to polaron calculations. We find 
$a = 2.49$~\AA\ and $c=6.47$~\AA\ for bulk h-BN, in good agreement with the experimental values 
$a = 2.50$~\AA\ and $c = 6.65$ \cite{Zhang2017}. For calculations of monolayer h-BN, we remove one BN 
layer from the bulk crystal structure, and we increase the $c$ parameter to $80$~\AA\ to minimize spurious 
interactions between periodic replicas \cite{Lebedev2016}. With this choice, monolayer h-BN exhibits a direct
band gap at $K$ as expected \cite{Elias2019}. The valence bands of monolayer h-BN are relatively 
insensitive to the size of the vacuum buffer (the choice of vacuum mainly affects the energetics of the interlayer 
state in the conduction bands at $\Gamma$). Our calculated Kohn-Sham band gaps of bulk and monolayer
h-BN are $E_g = 4.06$~eV and $4.69$~eV, respectively, and agree with previous calculations \cite{Ribeiro2011,Peng2012}.
In the case of bulk h-BN, we find in-plane and out-of-plane electronic dielectric constants 
$\epsilon_\parallel^\infty = 4.85$ and $\epsilon_\perp^\infty = 2.85$, respectively; and static dielectric
constant $\epsilon_\parallel^0 = 6.61$ and $\epsilon_\perp^0= 3.39$, respectively. These values are in
agreement with previous work \cite{Laturia2018}. We calculate a hole mass in monolayer h-BN of $0.65~m_e$, in good 
agreement with previous calculations yielding $0.61$-$0.82~m_e$ \cite{Qi2012,Ferreira2019,Drummond2020}.

For monolayer ZrS$_2$, we employ a kinetic energy cutoff of 125~Ry, a uniform and unshifted Brillouin 
zone grid of 12$\times$12$\times$2 points for ground-state calculations, and a grid of 12$\times$12$\times$1
points to compute phonons. The $c$-axis parameter is set to 60~\AA, and the optimized lattice parameter 
is $a=$~3.69~\AA, in good agreement with Ref.~\onlinecite{Zhang2015}. In the case of 
fluorinated graphene (CF), we use a planewaves cutoff of 105~Ry, a wavevector grid of 12$\times$12$\times$1
points for both ground-state structure and phonon calculations, and $c=$~80~\AA.
The optimized lattice parameter of monolayer CF is $a=$~2.56~\AA, in good agreement with Ref.~\onlinecite{Ciraci2011}. 
For monolayer BeO, we employ a cutoff of 125~Ry, a wavevector grid of 12$\times$12$\times$1 points for both
ground-state and phonon calculations,
and $c=$~80~\AA. The optimized lattice parameter of monolayer BeO is $a=$~2.68~\AA,
in line with Ref.~\onlinecite{Zhang2021}.

\raggedbottom

Calculations of polarons for monolayer h-BN, ZrS$_2$, fluorographene, and monolayer BeO
are performed using a slab/vacuum superlattice in a supercell geometry.
The long-range \F\ component of the electron-phonon matrix element is calculated following
Ref.~\onlinecite{Sio2022} as:
\begin{eqnarray} \label{eq:keyexpression}
 g^{\mathcal{L}}_{mn\nu}(\bk,\bq) &=& \frac{e^2}{2\ve_0  \Omega} (\hbar/2 \w_{\bq\nu})^{1/2}\!\!  
\sum_{\bG\ne-\bq} \frac{\<u_{m\bk+\bq+\bG}| u_{n\bk}\>}{\left|\bqin\!+\!\bGin\right|} 
\sum_\k M_\k^{-1/2} e^{-i(\bqin+\bGin) \cdot \btau_{\k \parallel}} 
\sum_{\a\b}Z^*_{\k,\a\b} e_{\k\b,\nu}(\bq) \nonumber \\ &\times&\left[\d_{\a,\parallel}\, i 
(\bq\!+\!\bG)_\a K(\bq\!+\!\bG,\tau_{\k z}) - \d_{\a,z}\, \frac{\D K(\bq\!+\!\bG,\tau_{\k z}) }{\D 
\tau_{\k z}} \right], \label{eq:frohlichquasi2d}
\end{eqnarray}
where $e$, $\ve_0$ and $\Omega$ are the electron charge, the dielectric permittivity of vacuum, and the
supercell volume, respectively. $\omega_{\bq\nu}$ is the frequency of a phonon with wavevector $\bq$
and branch index $\nu$, and $u_{n\bk}$ is the periodic part of the Bloch wavefunction for the wavevector
$\bk$ and band index $n$. $M_\k$ is the mass of atom $\kappa$ with equilibrium position $\btau_\kappa$
and Born effective charge tensor $Z_{\kappa,\a\b}^*$. $e_{\k\b,\nu}(\bq)$ indicates the vibrational
eigendisplacement, and the $\bG$'s are reciprocal lattice vectors of the supercell. The subscript $\parallel$
indicates in-plane directions, $z$ indicates the out-of-plane direction. 
The kernel function $K$ appearing in Eq.~\eqref{eq:keyexpression} is defined as:\\
\begin{eqnarray} \label{eq:kernel}
&& K(\bQ,\tau_z) = \frac{1}{\ggamma^- - \ggamma^+} \frac{1}{Q^2} \nonumber \\ && \times 
\Bigg\{\left[(\aalpha + \ggamma^- \bbeta)e^{\Qin \tau_z} + (\bbeta + \ggamma^-\aalpha)e^{-\Qin 
\tau_z}\right]\times \nonumber \\ && \times \Bigg[ (\aalpha + \ggamma^+ \bbeta) [e^{(\Qin-iQ_z) 
\tau_z} - e^{-(\Qin-iQ_z)d}](\Qin+iQ_z) - (\bbeta + \ggamma^+ \aalpha) [e^{-(\Qin+iQ_z) 
\tau_z}-e^{(\Qin+iQ_z) d}](\Qin-iQ_z) \nonumber \\ && + \frac{\aalpha + \ggamma^+ \bbeta}{e^{-iQ_z 
c+\eta}-1} [1-e^{-(\Qin-iQ_z) d}](\Qin+iQ_z) -\frac{\bbeta + \ggamma^+ \aalpha}{e^{-iQ_z c+\eta}-1} 
[1-e^{(\Qin+iQ_z) d}](\Qin-iQ_z) \nonumber \\ && + \frac{1}{e^{-iQ_z c+\eta}-1}  [e^{(\Qin-iQ_z) 
D}-1](\Qin+iQ_z) -\frac{\ggamma^+}{e^{-iQ_z c+\eta}-1}  [e^{-(\Qin+iQ_z) D}-1](\Qin-iQ_z) \Bigg] 
\nonumber \\ && + \left[(\aalpha + \ggamma^+ \bbeta)e^{\Qin \tau_z} + (\bbeta + \ggamma^+\aalpha)e^{-\Qin 
\tau_z}\right]\times \nonumber \\ && \times \Bigg[ (\aalpha + \ggamma^- \bbeta) [1 
-e^{(\Qin-iQ_z)\tau_z}](\Qin+iQ_z) - (\bbeta + \ggamma^- \aalpha) [1-e^{-(\Qin+iQ_z) 
\tau_z}](\Qin-iQ_z) \nonumber \\ && + \frac{\aalpha + \ggamma^- \bbeta}{e^{iQ_z c+\eta}-1} 
[1-e^{-(\Qin-iQ_z) d}](\Qin+iQ_z) - \frac{\bbeta + \ggamma^- \aalpha}{e^{iQ_z c+\eta}-1} 
[1-e^{(\Qin+iQ_z) d}](\Qin-iQ_z) \nonumber \\ && + [e^{(\Qin-iQ_z) D}-1](\Qin+iQ_z) - \ggamma^- 
[e^{-(\Qin+iQ_z) D}-1](\Qin-iQ_z) \nonumber \\ && + \frac{1}{e^{iQ_z c+\eta}-1} [e^{(\Qin-iQ_z) D}- 
1](\Qin+iQ_z) - \frac{\ggamma^-}{e^{iQ_z c+\eta}-1}   [e^{-(\Qin+iQ_z) D}-1](\Qin-iQ_z) \Bigg] 
\Bigg\}, \label{eq.fullkernel}
\end{eqnarray}
where the auxiliary quantities $\aalpha$, $\bbeta$, $\ggamma^{\pm}$, and $\eta$ appearing in these
expressions are defined as follows: 
\begin{eqnarray}
\aalpha &=& (1+1/\epsilon_{\infty})/2~, \label{eq.defs1}\\ \bbeta&=& 
(1-1/\epsilon_{\infty})/2~,\\ \ggamma^\pm &=& 
-\frac{ \displaystyle e^{\Qin D} - e^{\pm \eta}\left(\aalpha e^{-\Qin d} + \bbeta e^{\Qin d}\right)  
}{ \displaystyle e^{-\Qin D} - e^{\pm \eta}\left(\bbeta e^{-\Qin d} + \aalpha e^{\Qin d}\right)  
}~,\\ \eta &=& \cosh^{-1} \big\{ \cosh[\Qin(D-d)] + 2\aalpha^2/(2\aalpha - 1) 
\sinh(\Qin D) \sinh(\Qin d)  \big\}.\label{eq.eta} \phantom{\int} \label{eq.defs4}
\end{eqnarray}
In these expressions, $\epsilon_\infty$ is the electronic dielectric permittivity of the 2D layer, $d$ is the size 
of this layer, $D$ is the size of the vacuum buffer, and $c=d+D$ is the supercell size. $\epsilon_\infty$ and $d$
are defined unambiguously by requiring that the screening in the layer be isotropic, as originally proposed in
Ref.~\onlinecite{Sohier2016}. To this aim we use Eqs.~(64) and (65) of Ref.~\onlinecite{Sio2022}, and 
we obtain the effective thickness and dielectric constants shown in Supplemental Table~S1.

\begin{figure}
  \centering
  \includegraphics[width=\textwidth]{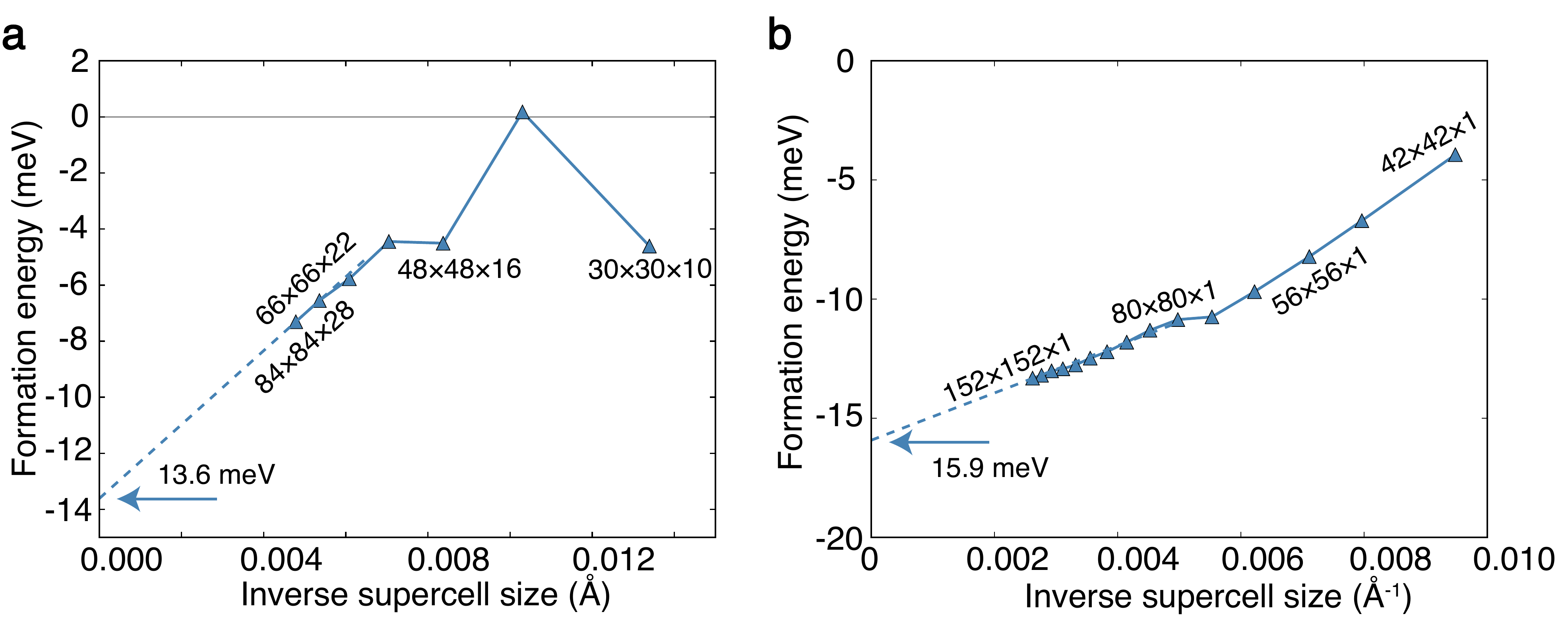}
  \caption{\label{fig.extrapolation}
  Formation energy of hole polaron in h-BN
  as a function of inverse supercell size, for (a) bulk h-BN and (b) monolayer h-BN. 
  The numbers next to some of the data points indicate the supercell dimension as a multiple of the crystalline unit
  cell. The dashed line is the linear extrapolation based on the last 3 data points, following the procedure outlined
  in Ref.~\onlinecite{Sio2019b}. The extrapolation to infinite supercell size provides the formation energy of the
  isolated polaron.}
\end{figure}

In Ref.~\onlinecite{Sio2022} it is shown that Eqs.~\eqref{eq:keyexpression}-\eqref{eq.defs4} provide
the exact long-wavelength limit of the electron-phonon matrix element in polar systems. These expressions
reduce to the standard \textit{ab initio} \F\ matrix element for bulk 3D systems \cite{Verdi2015,Sjakste2015},
and to the matrix elements for slabs in vacuum with Coulomb truncation for isolated 2D systems \cite{Sohier2016}.
The matrix element in Eq.~(3) of the main text was derived in Ref.~\onlinecite{Sio2022} starting from 
Eqs.~\eqref{eq:keyexpression}-\eqref{eq.defs4}. In our \textit{ab initio} calculations, we use $g^{\mathcal{L}}$ 
in conjunction with Wannier-Fourier interpolation \cite{Giustino2007}, as described in Refs.~\onlinecite{Verdi2015,Sio2022}.
To perform these calculations, we employ a supercell of size $c = 80$~\AA. We checked that the polaron formation
energy is well-converged with this vacuum buffer.
Using $c=80$~\AA\ we obtain a very small LO-TO splitting in monolayer h-BN, $\Delta \hbar\w =5$~meV.
This is consistent with the expectation that the splitting vanishes in the limit of infinite supercell size
\cite{Sohier2017}.
To correctly describe a single polaron in the entire crystal, we take the limit of infinite supercell
in the in-plane direction. Figure~\ref{fig.extrapolation}(a) shows the formation energy of polarons
in bulk h-BN for supercell sizes ranging from $30\times 30\times 10$ to $84\times 84 \times 28$. 
Figure~\ref{fig.extrapolation}(b) reports the corresponding data for monolayer h-BN, with supercell sizes 
ranging from $42\times 42\times 1$ to $152 \times 152\times 1$. 
The corresponding plots for monolayer ZrS$_2$, CF, and BeO are shown in Figs.~\ref{fig.extrapol-zrs2},
\ref{fig.extrapol-cf}, and \ref{fig.extrapol-beo}, respectively.

We note that the \textit{ab initio} polaron equations, Eqs.~(1)-(2) of the main text, provide a lower bound
to the polaron formation energy. In the case of 3D systems, the formation energy given by Eqs.~(1)-(2) 
is typically very accurate at strong coupling, and tends to underestimate the exact many-body energy
at weak coupling. This behavior is discussed in detail in Refs.~\onlinecite{Lafuente2022a, Lafuente2022b}.

\section*{Supplemental Note 1: Analysis of self-interaction error}

In this Note we discuss the impact of the self-interaction error in DFT calculations of polarons.
To keep the analysis as simple as possible, we consider the Landau-Pekar model of polarons in 3D,
see for example Refs.~\onlinecite{Devreese2020,Sio2019b}. This model describes an excess electron in 
a dielectric continuum, whose energy is given by:
  \begin{equation}
   E_{\rm LP} = 
  \frac{\hbar^2}{2m^*} \!\int\! d{\bf r}\, |\nabla \psi({\bf r})|^2 - \frac{e^2}{4\pi\varepsilon_0} 
  \left(\!\frac{1}{\epsilon_\infty}\!-\!\frac{1}{\epsilon_0}\!\right) \int \! d{\bf r}^\prime 
  \frac{\,\,|\psi({\bf r}^\prime)|^2}{|{\bf r}-{\bf r}^\prime|}.
  \end{equation}
Here, $\psi$ is the electron wavefunction and all other quantities are defined in the main text.
If we were to study this model within DFT, the electron would experience self-interaction. Considering
only the Hartree self-interaction for simplicity, the above energy would be modified as:
  \begin{equation}
  E_{\rm SI} = E_{\rm LP} + 
  \frac{e^2}{4\pi\varepsilon_0} \int \! d{\bf r}^\prime
  \frac{\,\,|\psi({\bf r}^\prime)|^2}{|{\bf r}-{\bf r}^\prime|}.
  \end{equation}
The Hartree self-interaction term and the electron-lattice attraction share the same integral,
therefore their prefactors can be combined as follows:
  \begin{equation}\label{eq.ESI}
   E_{\rm SI} = 
  \frac{\hbar^2}{2m^*} \!\int\! d{\bf r}\, |\nabla \psi({\bf r})|^2 + \frac{e^2}{4\pi\varepsilon_0} 
  \left(1-\!\frac{1}{\epsilon_\infty}\!+\!\frac{1}{\epsilon_0}\!\right) \int \! d{\bf r}^\prime 
  \frac{\,\,|\psi({\bf r}^\prime)|^2}{|{\bf r}-{\bf r}^\prime|}.
  \end{equation}
Since $\epsilon_0\ge \epsilon_\infty \ge 1$ for all materials, the quantity $(1-1/\epsilon_\infty+1/\epsilon_0)$ 
belongs to the interval $[0,1]$, hence the potential energy is repulsive. As a result, the energy minimum
is achieved for a fully-delocalized electron, and self-interaction has completely suppressed the formation of the
polaron.

Hubbard-corrected DFT and hybrid-functional DFT mitigate the self-interaction error of DFT, but do not
remove it completely. To see this, we rewrite Eq.~\eqref{eq.ESI} by adding an exchange term as in hybrid-functional DFT: 
  \begin{equation}
   E_{\rm Hy} = E_{\rm SI} - 
  \alpha \frac{e^2}{4\pi\varepsilon_0} \int \! d{\bf r}^\prime 
  \frac{\,\,|\psi({\bf r}^\prime)|^2}{|{\bf r}-{\bf r}^\prime|},
  \end{equation}
where $\alpha$ is the fraction of exact exchange in the hybrid functional.
With this correction term, polaron localization does not occur for $\alpha < \alpha_{\rm c}$, where
the critical fraction of exact exchange is $\alpha_{\rm c} = 1 - 1/\epsilon_\infty+1/\epsilon_0$.
For $\alpha > \alpha_{\rm c}$, a polaron forms, but its formation energy increases quadratically
with $\alpha$, $E_{\rm Hy} = -\mbox{const} \cdot (\alpha-\alpha_{\rm c})^2$.
These trends are a manifestation of residual self-interaction.

This analysis is consistent with the hybrid
functional calculations of Ref.~\onlinecite{kokott2018}: in that
work the authors find small polarons in MgO only for $\alpha > 0.48$, and the polaron formation energy
increases with $\alpha$ beyond this critical value. Our present estimate 
(with $\epsilon_\infty=2.4$ and $\epsilon_0=9.8$ for MgO), yields $\alpha_{\rm c}= 0.68$,
which is in good agreement with Ref.~\onlinecite{kokott2018} considering 
that our estimate refers to the limit of infinite supercell, whereas the calculation of 
Ref.~\onlinecite{kokott2018} is for a 3$\times$3$\times$3 supercell.

Unlike DFT, Hubbard-corrected DFT, and hybrid-functional DFT, 
the self-interaction error is not present in Eqs.~(1)-(2) of the main text, since these equations were
derived in Ref.~\onlinecite{Sio2019b} starting from a self-interaction-corrected functional.

\section*{Supplemental Note 2: Uncertainty quantification for h-BN}

The calculations for bulk h-BN presented in this manuscript employ band structures, phonon dispersions, and electron-phonon
matrix elements obtained from DFT/LDA. We expect that the use of non-local functionals such as hybrid functionals
or the recently-developed GW perturbation theory (GWPT) \cite{louie2019}
will modify the polaron formation energy to some extent, but it will not affect our conclusions, as shown in the following.

Hybrid-functional calculations or GWPT are currently beyond reach for the system under consideration,
which requires a supercell with 36,000 atoms. Nevertheless, we can estimate the correction to the formation 
energy using the 3D Landau-Pekar model introduced in Supplemental Note~1 and discussed in more detail in 
Supplemental Note~5 \cite{Devreese2020,Sio2019b}. In this model, the formation energy of the polaron scales with the 3D
Fr\"ohlich coupling constant $\alpha_{\rm F}$ \cite{Devreese2020}:
  \begin{equation}\label{eq.alphaF}
     \alpha_{\rm F} = \frac{e^2}{4\pi\varepsilon_0\hbar}
\left(\frac{1}{\epsilon_\infty}-\frac{1}{\epsilon_0}\right)
\sqrt{\frac{m^*}{2\hbar\omega_{\text{LO}}}}.
  \end{equation}
The materials parameters entering this expression are the effective mass $m^*$, the frequency of longitudinal-optical
phonons $\omega_{\rm LO}$, the high-frequency dielectric constant $\epsilon_\infty$, and the static dielectric constant 
$\epsilon_0$. Our DFT/LDA calculations yield the following values:
  $$
  m^* = 0.54~m_e, \qquad \omega_{\rm LO} = 201~\mbox{meV}, \qquad
   \varepsilon_\infty = 4.85, \qquad \varepsilon_0 = 6.61.
  $$
Using these values in Eq.~\eqref{eq.alphaF}, we obtain $\alpha_{\rm F} = 0.331$.
If we evaluate instead the coupling constant using experimental values for the same parameters,
namely the effective mass from Ref.~\onlinecite{Henck2017} and all other
parameters from Ref.~\onlinecite{Geick1966}:
  $$
  m^* = 0.49~m_e, \qquad \omega_{\rm LO} = 200~\mbox{meV}, \qquad
   \varepsilon_\infty = 4.95, \qquad \varepsilon_0 = 6.85,
  $$
we obtain $\alpha_{\rm F}^{\rm exp} = 0.323$. This value is within 3\% of our DFT/LDA
result, therefore we expect a negligible change in the polaron formation energy.
In fact, in the weak-coupling limit, the energy is given by $\alpha_{\rm F}\hbar\omega_{\rm LO}$, while
in the strong coupling limit it is given by $\alpha_{\rm F}^2\hbar\omega_{\rm LO}/3\pi$ \cite{Devreese2020}. 
In either case, the corrected energy is within 5\% of our original result, thus supporting our initial choice of using
DFT/LDA for h-BN.

\section*{Supplemental Note 3: Derivation of Eq.~(4) of the main text}

In this Note we outline the key steps of the derivation of the 
polaron formation energy in two dimensions, as given in Eq.~(5) of the main text. 
We refer to Ref.~\cite{Sio2019a,Sio2019b} for a derivation of the 
\textit{ab initio} polaron equations, Eqs.~(2) and (3) of the main text.

We begin by rewriting Eqs.~(2) and (3) of the main text for the
case of a single parabolic electron band with effective mass $m^*$, and a single dispersionless longitudinal 
optical phonon with frequency $\w$ (the discussion of holes leads to the same results): 
 \begin{eqnarray}
 && \frac{2}{\Omega_{\rm BZ}}\int_{\rm BZ} d\bq \, B_{\bq}\,g^*_\bq\,A_{\bk+\bq}=(\ve_{\bk}-\ve)\,A_{\bk}, 
     \label{eq.si.1}\\
 && B_{\bq}= \frac{g_\bq}{\hbar\w} \frac{1}{\Omega_{\rm BZ}}\int_{\rm BZ} d\bk \, A^*_{\bk+\bq} A_{\bk}.
     \label{eq.si.2}
 \end{eqnarray}
In these expressions, $\bk$ and $\bq$ are electron and phonon wavevectors, respectively, 
$g_\bq$ is the \F\ matrix element, $\Omega_{\rm BZ}$ is the volume of the Brillouin zone, and the limit of dense
Brillouin zone sampling has been taken. $A_\bk$ and $B_\bq$ are the amplitudes of the polaron wavefunctions
and associated normal mode coordinates, and $\ve$ is the polaron energy eigenvalue. The electron bands
are given by $\ve_\bk = \hbar^2 |\bk|^2/2m^*$. 
 These expressions are for a standard three-dimensional (3D) bulk crystal,
as in Ref.~\cite{Sio2019a,Sio2019b}.

Using the same assumptions of single band and single LO phonon, the formation energy of the polaron
given in Eq.~(42) of Ref.~\cite{Sio2019b} can be written as follows:
  \begin{equation}
     E = \ve + \hbar\w \int_{\rm BZ}\frac{d\bq}{\Omega_{\rm BZ}} |B_\bq|^2,
     \label{eq.si.3}
  \end{equation}
where we have taken the zero of the energy to coincide with the conduction band bottom.
By combining Eqs.~\eqref{eq.si.1}-\eqref{eq.si.3} we obtain:
  \begin{equation}\label{eq.f.en}
     E = \frac{1}{\Omega_{\rm BZ}}\int_{\rm BZ} d\bk \, \epsilon_\bk |A_{\bk}|^2 
       - \frac{1}{\hbar\w} \int_{\rm BZ}\frac{d\bq}{\Omega_{\rm BZ}} |g_\bq|^2 |f_\bq|^2,
  \end{equation}
where we have used the normalization of the polaron amplitudes \cite{Sio2019b}:
  \begin{equation}
 \frac{1}{\Omega_{\rm BZ}}\int_{\rm BZ} \!\!d\bk \, |A_{\bk}|^2 = 1,
  \end{equation}
and we have introduced the auxiliary function:
  \begin{equation}\label{eq.aux}
  f_{\bq}=  \frac{1}{\Omega_{\rm BZ}}\int_{\rm BZ} \!\!d\bk \, A^*_{\bk+\bq} A_{\bk}.
  \end{equation}
The polaron wavefunction is given by Eq.~(33) of Ref.~\cite{Sio2019b}, which we rewrite here as:
  \begin{equation}\label{eq.wfc1}
  \psi(\br) = \frac{1}{\Omega_{\rm BZ}}\int_{\rm BZ}d\bk\, A_\bk u_\bk(\br) e^{i\bk\cdot\br}.
  \end{equation}
In this expression, $u_\bk$ is the Bloch-periodic component of the wavefunction, and $\br$ is the
position vector in 3D. We now specialize to the case of a 2D material of thickness $d$, contained
in an otherwise empty supercell of size $L$. We describe the confinement of the electronic states
in the slab using a simple square function:
  \begin{equation}\label{eq.profile}
     u_\bk(\br) = \begin{cases} (Ad)^{-1/2} & 0<z<d \\ 0 & d<z<L \end{cases},
  \end{equation}
where $A$ is the area of the unit cell in the plane of the slab. Other choices for the profile of
the wavefunction along the $z$ axis lead to the same results in the limit $L \gg d$.
When we take the limit $L \gg d$, the Brillouin zone size along the $z$ axis tends to vanish,
therefore $A_\bk$ can be replaced by $A_{k_x,k_y,0}$. We will indicate this quantity as $A_{\bk_\parallel}$.
Similarly we will use $\br_\parallel$ to indicate the in-plane position vector, and $\Omega_{\rm BZ_\parallel}=
(2\pi)^2/A$ to denote the 2D Brillouin zone. With the choice
made in Eq.~\eqref{eq.profile}, the wavefunction in Eq.~\eqref{eq.wfc1} can be rewritten as:
  \begin{equation}
  \psi(\br) = \int_{\rm BZ_\parallel}\frac{d\bk_\parallel}{\Omega_{\rm BZ_\parallel}} 
     \, A_{\bk_\parallel} e^{i\bk_\parallel\cdot\br_\parallel}
       \times \begin{cases}
     (Ad)^{-1/2} {\rm sinc} (\pi z/L)e^{i\pi z/L} & 0<z<d \\ 0 & d<z<L \end{cases}.
  \end{equation}
In the limit $L \gg d$, the terms dependent on $z$ approach unity as $z$ is bound by $d$, therefore the
last expression simplifies to:
  \begin{equation}
  \psi(\br_\parallel,z) = \frac{1}{\sqrt{A}}
       \int_{\rm BZ_\parallel}\frac{d\bk_\parallel}{\Omega_{\rm BZ_\parallel}} 
      \, A_{\bk_\parallel} e^{i\bk_\parallel\cdot\br_\parallel}
       \times \begin{cases}
     1/\sqrt{d} & 0<z<d \\ 0 & d<z<L \end{cases}.
  \end{equation}
Let us call $\phi(\br_\parallel)$ the first term in the product on the right-hand side of the last equation.
We have:
  \begin{equation}\label{eq.phi}
  \phi(\br_\parallel) = \frac{1}{\sqrt{A}}
       \int_{\rm BZ_\parallel}\frac{d\bk_\parallel}{\Omega_{\rm BZ_\parallel}} 
      \, A_{\bk_\parallel} e^{i\bk_\parallel\cdot\br_\parallel}.
  \end{equation}
This relation can be inverted to obtain $A_{\bk_\parallel}$ in terms of $\phi(\br_\parallel)$:
  \begin{equation}\label{eq.A}
   A_{\bk_\parallel}  = \frac{1}{\sqrt{A}} \int \!d\br_\parallel \,e^{-i\bk_\parallel\cdot \br_\parallel} \phi(\br_\parallel).
  \end{equation}
% To do this I wrote S11 using a discrete set of kpoints and then taken the limit of dense sampling at the end.
% Units OK.
We now replace this expression inside Eq.~\eqref{eq.aux}. After using Eq.~\eqref{eq.phi} we find:
%  \begin{equation}
%  f_{\bq}=  \frac{1}{A_{\rm BZ}}\int_{\rm BZ_\parallel} \!\!d\bk_\parallel \, A^*_{\bk+\bq} 
%      \frac{1}{\sqrt{A}} \int \!d\br_\parallel \,e^{-i\bk_\parallel\cdot \br_\parallel} \phi(\br_\parallel).
%  \end{equation}
  \begin{equation}\label{eq.aux2}
  f_{\bq}=  \!\int \!\!d\br_\parallel \,\, e^{i\bq\cdot \br_\parallel} |\phi(\br_\parallel)|^2 .
  \end{equation}
% Units OK.
Using this result, the second term on the right-hand side of Eq.~\eqref{eq.f.en} becomes:
  \begin{equation}\label{eq.pot}
       - \frac{1}{\hbar\w} \int_{\rm BZ}\frac{d\bq}{\Omega_{\rm BZ}} |g_\bq|^2 |f_\bq|^2
      = 
     \frac{1}{2}\!\int \!\!d\br_\parallel \,d\br_\parallel' \,
    |\phi(\br_\parallel)|^2
         V_{\rm 2D}(\br_\parallel-\br_\parallel')
    |\phi(\br_\parallel')|^2,
  \end{equation}
having defined:
  \begin{equation}\label{eq.V2d}
  V_{\rm 2D}(\br_\parallel) = - \frac{2}{\hbar\w} \int_{\rm BZ_\parallel} \!\frac{d\bq_\parallel}{\Omega_{\rm BZ_\parallel}} 
     \, e^{i\bq\cdot \br_\parallel} |g_\bq|^2 .
  \end{equation}
The first term on the right-hand side of Eq.~\eqref{eq.f.en} can be obtained along similar steps. 
First, we replace the expression for $A_\bk$ from
Eq.~\eqref{eq.A}:
  \begin{equation}
     \frac{1}{\Omega_{\rm BZ}}\int_{\rm BZ} d\bk \, \epsilon_\bk |A_{\bk}|^2 
        = \frac{1}{\sqrt{A}}
           \!\int \!\!d\br_\parallel \phi(\br_\parallel)   
   \int_{\rm BZ_\parallel} \!\frac{d\bk_\parallel}{\Omega_{\rm BZ_\parallel}} 
          A_\bk^*  \frac{\hbar^2|\bk|^2}{2m^*} e^{-i\bk\cdot \br_\parallel}.
  \end{equation}
Then, we rewrite the Brillouin-zone integral on the right using the real-space kinetic energy operator,
and we substitute Eq.~\eqref{eq.phi}:
  \begin{equation}\label{eq.kin}
     \frac{1}{\Omega_{\rm BZ}}\int_{\rm BZ} d\bk \, \epsilon_\bk |A_{\bk}|^2 
        = -\!\int \!\!d\br_\parallel \phi(\br_\parallel)   
        \frac{\hbar^2\nabla^2}{2m^*} \phi^*(\br_\parallel).
  \end{equation}
By combining Eqs.~\eqref{eq.kin} and \eqref{eq.pot} inside Eq.~\eqref{eq.f.en}, we obtain Eq.~(4) of the
main text.
   
\section*{Supplemental Note 4: Derivation of Eqs.~(5) and (6) of the main text}

In this Note we derive the effective self-trapping potential $V_{\rm 2D}$ in Eqs.~(5) and (6) of the main text.
To this end, we consider the expression for the \F\ matrix element in 2D as derived in Eq.~(62) of 
Ref.~\cite{Sio2022}, and reproduced in the main text as Eq.~(3):
 \begin{equation}\label{eq.gg}
   g(q) = \sqrt{E_{\rm Ha}\frac{\pi}{2} \frac{a_0 d}{A} \, \hbar\w\, \epsilon_{\rm ion}}
   \, \frac{q_c}{q_c+q}.
 \end{equation}
In this expression, $q=|\bq_\parallel|$, $E_{\rm Ha}$ is the Hartree energy, $a_0$ is the Bohr radius,
and $\epsilon_{\rm ion}$ is the ionic contribution to the dielectric permittivity. $q_c$ is a characteristic
wavevector defined in the main text. We replace Eq.~\eqref{eq.gg} inside Eq.~\eqref{eq.V2d} to obtain:
  \begin{equation}
  V_{\rm 2D}(\br_\parallel) = -  E_{\rm Ha} \frac{\pi a_0 d}{(2\pi)^2} \,\epsilon_{\rm ion}\, q_c^2
     \int_{\rm BZ_\parallel} 
   \frac{e^{i\bq\cdot \br_\parallel} d\bq_\parallel}{(q_c+q)^2}.
  \end{equation}
The integral can be carried out in cylindrical coordinates, after replacing the Brillouin zone area by
an equivalent disk of radius $\pi q_{\rm BZ_\parallel}^2 = \Omega_{\rm BZ_\parallel}$:
  \begin{equation}\label{eq.tmp.v2d}
  V_{\rm 2D}(\br_\parallel) = -  E_{\rm Ha} \frac{\pi a_0 d}{(2\pi)^2} \,\epsilon_{\rm ion}\, q_c^2
     \int_0^{q_{\rm BZ_\parallel}} 
   \frac{q \,dq}{(q_c+q)^2}  \int_0^{2\pi} d\theta e^{i q r \cos\theta}.
  \end{equation}
In this expression, $r=|\br_\parallel|$. The angular integral on the right 
is $2\pi J_0(qr)$, with $J_0$ denoting the Bessel function of the first kind. Therefore we have:
  \begin{equation}\label{eq.bound1}
  V_{\rm 2D}(\br_\parallel) = -  E_{\rm Ha} \frac{a_0 d}{2} \,\epsilon_{\rm ion}\, q_c^2
     \int_0^{q_{\rm BZ_\parallel}r} 
   \frac{x J_0(x) \,dx}{(q_c r +x)^2}.
  \end{equation}
The integration bound can be extended to infinity without making a significant error, as it is usually
done for the \F\ model in 3D systems. Using this approximation we obtain Eq.~(5) of the main text:
% This should be OK when $q_cr$ in the denominator is small.  Since $q_c r \sim q_{BZ} r$, when $q_c r$ 
% is small we can probably extend to infinity. When it is large, we have that the $q_{BZ} r$ is large anyway. 
  \begin{equation}
  V_{\rm 2D}(\br_\parallel) = -  E_{\rm Ha} \frac{a_0 d}{2} \,\epsilon_{\rm ion}\, q_c^2
       \phi(q_c r),
  \end{equation}
where the function $\phi$ is defined as:
  \begin{equation}
  \phi(s) = \int_0^\infty \frac{x J_0(x) \,dx}{(s +x)^2}.
  \end{equation}
To obtain Eq.~(6) of the main text we need to carry out the integration in the last expression.
We recast this expression as follows:
%  \begin{equation}
%  \phi(s) =  - \frac{d}{ds} \int_0^\infty \frac{x J_0(x)}{x+s} 
%  \end{equation}
%  $$
%  \phi(s) = - \frac{d}{ds} \int_0^\infty \frac{(x+s-s) J_0(x)}{x+s} 
%  $$
%  $$
%  \phi(s) = - \frac{d}{ds} \int_0^\infty J_0(x) dx
%   + \frac{d}{ds} s \int_0^\infty \frac{J_0(x)}{x+s} dx
%   = \left[1+ s \frac{d}{ds} \right] \int_0^\infty \frac{J_0(x)}{x+s} dx
%  $$
  \begin{equation}
  \phi(s) =  \left[1+ s \frac{d}{ds} \right] \int_0^\infty \frac{J_0(x)}{x+s} dx.
  \end{equation}
The last integral is equal to $\pi[H_0(s)-Y_0(s)]/2$, 
% from Mathematica. This could be done using the laplace transform of J0 but it does not seem worth the effort
where $H_0$ is the Struve function and
$Y_0$ is the Bessel function of the second kind. After carrying our the derivative in the last
equation, we find the Struve and Bessel functions $H_1$ and $Y_1$, leading to Eq.~(6) of the main text:
  \begin{equation}
    \phi(s) = s + \frac{\pi}{2}\left[ H_0(s)-Y_0(s)\right] - \frac{\pi}{2} s\left[  H_1(s)- Y_1(s)\right].
  \end{equation}
  
\section*{Supplemental Note 5: Derivation of Eq.~(7) of the main text}

In this Note we derive Eq.~(7) of the main text, which gives the formation energy of the polaron
corresponding to the variational ansatz: 
 \begin{equation}\label{eq.ans}
    \phi(\br_\parallel) = \sqrt{\frac{2}{\pi r_p^2}} \exp(-r/r_p).
 \end{equation}
For convenience we reproduce below the formation energy as obtained by combining Eqs.~\eqref{eq.f.en},
\eqref{eq.pot}, \eqref{eq.V2d}, \eqref{eq.kin}, and \eqref{eq.gg}:
\begin{equation}\label{eq.ef.1}
     E = \frac{\hbar^2}{2m^*} \!\int \!\!d\br_\parallel |\nabla\phi(\br_\parallel)|^2 - 
      \frac{e^2 \epsilon_{\rm ion}\, q_c^2 d}{16\pi \ve_0} \frac{1}{2\pi} 
       \int_{\rm BZ_\parallel} \! d\bq_\parallel\,
       \frac{1}{(q_c+q)^2}
     \left|\int \!\!d\br_\parallel e^{i\bq\cdot \br_\parallel} |\phi(\br_\parallel)|^2  \right|^2.
\end{equation}
Replacing the ansatz of Eq.~\eqref{eq.ans} in the first term of Eq.~\eqref{eq.ef.1} gives:
  \begin{equation}\label{eq.kin.fin}
      \frac{\hbar^2}{2m^*} \!\int \!\!d\br_\parallel |\nabla\phi^*(\br_\parallel)|^2  = 
      \frac{\hbar^2}{2m^*} \frac{2}{\pi r_p^4} \int \!\!d\br_\parallel  e^{-2 r/r_p}
      = \frac{\hbar^2}{2m^* r_p^2}.
  \end{equation}
This is the kinetic term in Eq.~(7) of the main text. For the second term of the right-hand side of 
Eq.~\eqref{eq.ef.1}, we first evaluate the Fourier transform of $|\phi(\br_\parallel)|^2$:
  \begin{equation}
    \int \!\!d\br_\parallel e^{i\bq\cdot \br_\parallel} |\phi(\br_\parallel)|^2 =
    \frac{2}{\pi r_p^2} \int_0^\infty \!\!\!\!dr\, r  e^{-2 r/r_p}  \int_0^{2\pi} \!\!d\theta\, e^{i q r \cos\theta} .
  \end{equation}
As in Eq.~\eqref{eq.tmp.v2d}, the last integral is $2\pi J_0(qr)$, therefore we have:
  \begin{equation}
    \int \!\!d\br_\parallel e^{i\bq\cdot \br_\parallel} |\phi(\br_\parallel)|^2 =
    s^2 \int_0^\infty \!\!\!\!d x \, x \, e^{-s x}  J_0(x)
    = -s^2 \frac{d}{d s}\int_0^\infty \!\!\!\!d x \,  e^{-s x}  J_0(x), \qquad s=\frac{2}{q r_p}.
  \end{equation}
The integral on the right is the Laplace transform of the Bessel function, $(1+s^2)^{-1/2}$, therefore we
can write: 
  \begin{equation}
    \int \!\!d\br_\parallel e^{i\bq\cdot \br_\parallel} |\phi(\br_\parallel)|^2 = 8(4+q^2r_p^2)^{-3/2}.
  \end{equation}
Using this result in the second term of Eq.~\eqref{eq.ef.1} we obtain:
  \begin{equation}\label{eq.pot.fin}
     - \frac{e^2 \epsilon_{\rm ion}\, q_c^2 d}{16\pi \ve_0} \frac{1}{2\pi} 
       \int_{\rm BZ_\parallel} \! d\bq_\parallel\, \frac{1}{(q_c+q)^2}
     \left|\int \!\!d\br_\parallel e^{i\bq\cdot \br_\parallel} |\phi(\br_\parallel)|^2  \right|^2
      = 
     - \frac{e^2 \epsilon_{\rm ion}\, q_c^2 d}{16\pi \ve_0} \,64\!\!
       \int_0^{q_{\rm BZ_\parallel} r_p} \!\!d x\,\frac{x}{(q_c r_p + x)^2(4+x^2)^3}.
\end{equation}
As for Eq.~\eqref{eq.bound1}, we can extend the integration bound to infinity without making a significant
error. Putting Eqs.~\eqref{eq.kin.fin} and \eqref{eq.pot.fin} together, we find Eq.~(7) of the main text:
   \begin{equation}\label{eq.ef}
     E = \frac{\hbar^2}{2m^* r_p^2} 
     - \frac{\epsilon_{\rm ion}}{4}\frac{e^2 \, d}{4\pi \ve_0 r_p^2}  f(q_c r_p),
   \end{equation} 
where the function $f$ is given by:
   \begin{equation}\label{eq.chi}
       f(s) = 64 \,s^2\!\!\int_0^\infty \!\!d x\,\frac{x}{(s + x)^2(4+x^2)^3}.
   \end{equation} 
Integration with Mathematica gives the following 
explicit expression for $f$:
 \begin{equation}\label{eq.chi2}
  f(s) = \frac{s^2}{(s^2+4)^4} \left[ s^6 -\pi s^5 +28s^4 -40 \pi s^3
 - 16 (1+20\log 2) s^2 +240\pi s + 64(4 \log 2 -7) +64 (5s^2-4)\log s \right].
 \end{equation}
A plot of this function is shown in Fig.~\ref{fig.f-function}.

\section*{Supplemental Note 6: Weak coupling expansion}

In this Note we derive the weak coupling expansion for the polaron formation energy $E$ and radius $r_p$
in terms of the descriptors $\beta$ and $\gamma$ defined in the main text:
  \begin{equation} \label{eq.beta}
     \beta = \epsilon_{\rm ion}\frac{m^*}{m_0}\frac{d}{a_0},
  \end{equation}
  \begin{equation} \label{eq.gamma}
     \gamma = \frac{(q_c a_0)^2}{m^*/m_0}, \qquad q_c = \frac{1}{d}\frac{4\epsilon_\infty}{2\epsilon_\infty^2-1}.
  \end{equation}
Using Eqs.~\eqref{eq.beta}-\eqref{eq.gamma}, Eq.~\eqref{eq.ef} can be rewritten as:
   \begin{equation}\label{eq.ef2}
     \frac{E}{E_{\rm Ha}} = \frac{\gamma }{4} \frac{2 - \beta  f(s)}{s^2}, \qquad s = q_c r_p,
   \end{equation} 
with $f$ given by Eq.~\eqref{eq.chi2}. The polaron formation energy $E$ is found by minimizing this
expression with respect to the scaled polaron radius $s = q_c r_p$. The location of this minimum
only depends on the parameter $\beta$, therefore the polaron radius $r_p$ depends on $\beta$ and $q_c$.
Similarly, the value of the formation energy at the minimum depends only on $\beta$ and $\gamma$.

The weak coupling case corresponds to the limit of $\beta$ approaching 2 from above. 
In this limit, the polaron radius
tends to infinity, therefore we can replace the function $f$ in Eq.~\eqref{eq.chi2} by its asymptotic
expansion at large $s$:
  \begin{equation}
   f(s) = 1-\frac{\pi}{s}, \quad s\gg 1.
  \end{equation}
A plot of this approximation is shown in Fig.~\ref{fig.f-function}. Using this expression inside 
Eq.~\eqref{eq.ef2}, and retaining only the leading-order terms in $\b-2$, we obtain the polaron radius
and energy in the weak-coupling limit:
  \begin{equation}
% exact minimum: s = \frac{3\pi\beta}{2(\beta-2)}, then take beta=2
    r_p = \frac{1}{q_c}\frac{3\pi}{\beta-2},
  \end{equation}
  \begin{equation}\label{eq.wc}
     \frac{E}{E_{\rm Ha}} = -\frac{1}{108\, \pi^2} \gamma (\b-2)^3.
  \end{equation}
Equation~\eqref{eq.wc} yields a formation energy that is within 10\% of the value obtained by direct
numerical minimization of Eq.~\eqref{eq.ef2} for $\beta < 2.5$.

\section*{Supplemental Note S7: Polarons in monolayer ZrS$_2$, fluorographene, and monolayer BeO}

In this Note we report additional calculations of polarons in 2D crystals. We use 
the \textit{ab initio} polaron equations reported as Eqs.~(1)-(2) of the main text, we perform
validation tests via self-interaction-corrected DFT (SIC-DFT) and hybrid functionals,
and we compare with the analytical model in Eq.~(7) of the main text. We consider
the following systems:
  \begin{itemize}\itemsep0pt
    \item[1.] Electron and hole polarons in monolayer ZrS$_2$;
    \item[2.] Electron and hole polarons in fluorographene (CF);
    \item[3.] Hole polaron in monolayer BeO.
  \end{itemize}

\textbf{1. Electron and hole polarons in monolayer ZrS$_\mathbf{2}$}

\vspace{2pt}
\textit{Calculations using the \textit{ab initio} polaron equations}

Our calculations indicate that monolayer ZrS$_2$ hosts large electron and hole polarons. The electron polaron is comprised 
primarily of Zr-$4d$ states as shown by the wavefunction plot in Fig.~\ref{fig.zrs2-wfc}(a). 
The wavefunction is spatially anisotropic, and we find three degenerate polarons which are related by 60$^\circ$ 
rotations. Each polaron is formed by Bloch states belonging to pairs of $M$ points that are time-reversal partners, 
as well as the conduction band minimum at $\Gamma$, as shown in Fig.~\ref{fig.zrs2-wfc}(b) and (c). 
The contribution of vibrational modes to the polaron formation 
energy are shown in Fig.~\ref{fig.zrs2-wfc}(d), and the corresponding density of states is in 
Fig.~\ref{fig.zrs2-wfc}(e). The formation energy of the electron polaron in the dilute limit is 189~meV. 

Figure~\ref{fig.zrs2-wfc-hole} shows the corresponding results for the hole polaron in monolayer ZrS$_2$.
In this case, the hole polaron draws weigth primarily from the top of the valence band at $\Gamma$, which
has S-$3p$ character. The formation energy of the hole polaron in the dilute limit is 110~meV.

\vspace{5pt}
\textit{Calculations using self-interaction corrected DFT}

\begin{figure}
  \centering
  \includegraphics[width=\textwidth]{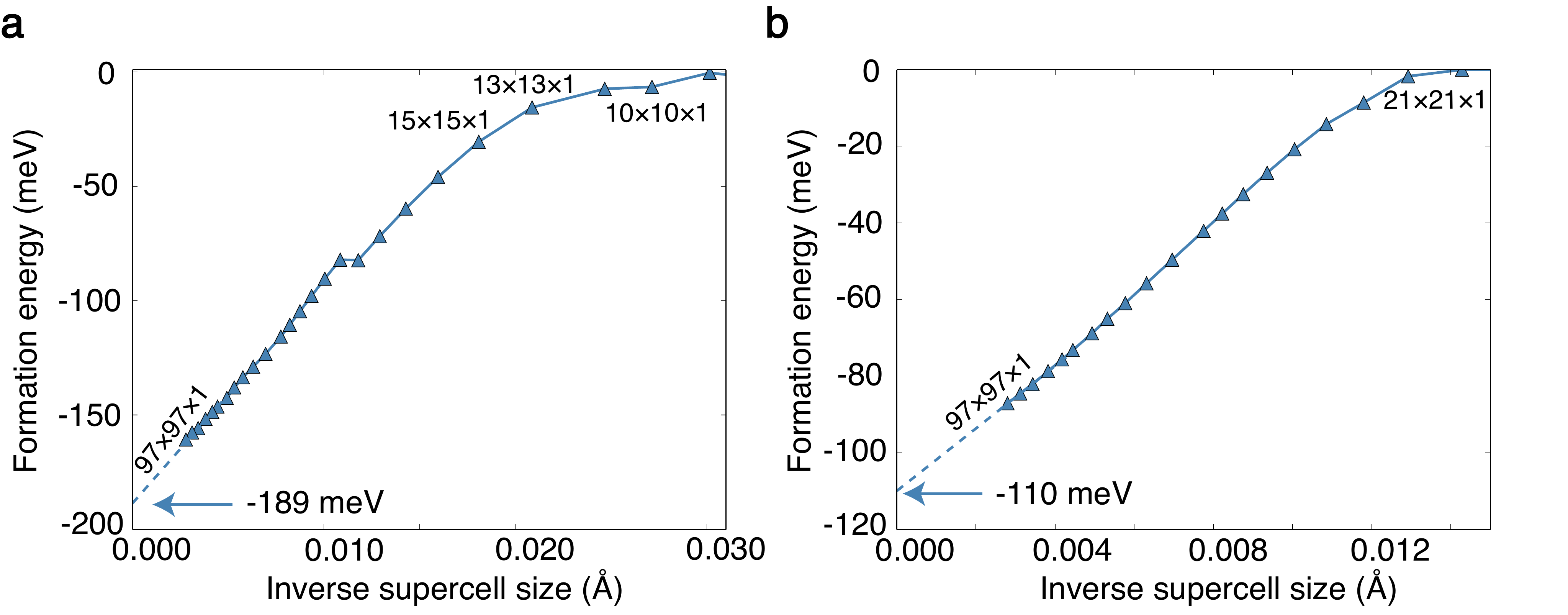}
  \caption{\label{fig.extrapol-zrs2}
  Formation energy of (a) electron polaron and (b) hole polaron of monolayer ZrS$_2$.
  The numbers next to some of the data points indicate the supercell dimension as a multiple of the crystalline unit
  cell. The dashed line is the linear extrapolation as in Fig.~\ref{fig.extrapolation}, and the intercept with the
  vertical axis indicates the formation energy in the dilute limit.
}
\end{figure}

To validate our results for ZrS$_2$ obtained from the \textit{ab initio} polaron equations, we perform explicit 
self-interaction-corrected (SIC) DFT calculations using the method of Ref.~\onlinecite{Sio2019b}.
We consider a supercell of 7$\times$7$\times$1 unit cells (147 atoms). This calculation
does not indicate any stable electron polaron, in agreement with the fact that polaron formation
is only observed with the \textit{ab initio} polaron equations [Fig.~\ref{fig.extrapol-zrs2}(a)] 
for supercells of at least 10$\times$10$\times$1 unit cells (300 atoms). 
Therefore, both the \textit{ab initio} polaron equations and SIC-DFT indicate that there are no 
small electron polarons in ZrS$_2$.

\vspace{5pt}
\textit{Calculations using the 2D model}

The model of Eq.~(7) yields a formation energy of 110~meV for the electron polaron. In this estimate
we employed the dielectric constants reported in Supplemental Table~S1.
In comparison, the \textit{ab initio} polaron equations yield
a formation energy of 189~meV, of which 64\% results from LO phonons [see Fig.~\ref{fig.zrs2-wfc}(e)]. 
Therefore the formation energy arising only from Fr\"ohlich interactions 
is 121~meV, in good agreement with the 2D model.

The model of Eq.~(7) also predicts a formation energy of 104~meV for the hole polaron. In comparison,
the \textit{ab initio} polaron equations yield a formation energy of 110~meV, of which 94\% results 
from LO phonons [see Fig.~\ref{fig.zrs2-wfc-hole}(d)]. Therefore the contribution
of Fr\"ohlich interactions to the formation energy is of 103~meV, which is also in good agreement with the model. 

We note that the formation energy is rather sensitive to the values of the dielectric constants.
For example, by using the values reported in Ref.~\onlinecite{Pike2018} 
($\varepsilon_\infty = 9.9$, $\varepsilon_0 = 37.2$), instead of those reported in
Supplemental Table~S1 ($\varepsilon_\infty = 12.2$, $\varepsilon_0 = 29.6$), we obtain formation energies
of 349~meV and 313~meV for the electron and hole polaron, respectively, as shown in Table~S3. This is because
the formation energy is rather sensitive to $\epsilon_{\rm ion}$ via the descriptor $\beta$, see Eq.~\eqref{eq.wc}.
Therefore care must be used
when estimating polaron formation energies in 2D materials using dielectric constants from bulk 3D compounds.

\vspace{5pt}
\textbf{2. Electron and hole polarons in fluorographene}

\vspace{2pt}
\textit{Calculations using the \textit{ab initio} polaron equations}

\begin{figure}
  \centering
  \includegraphics[width=\textwidth]{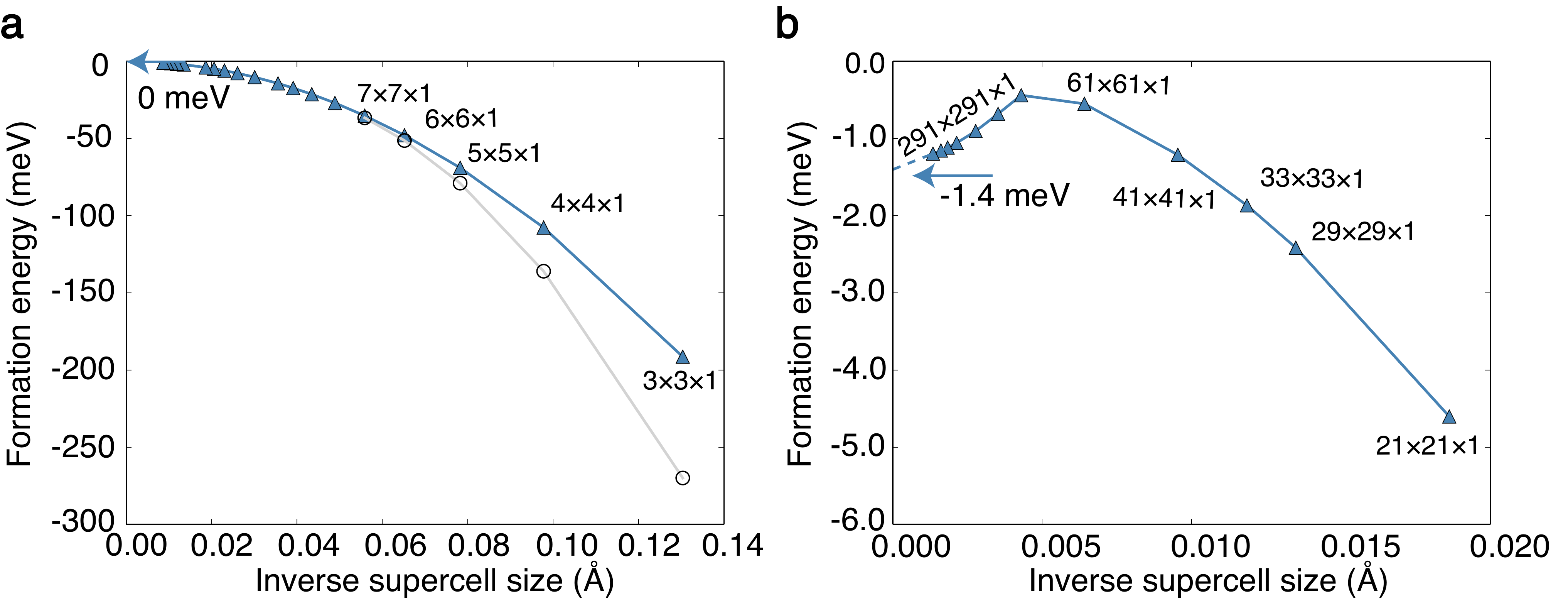}
  \caption{\label{fig.extrapol-cf}
  Formation energy of (a) electron polaron and (b) hole polaron of monolayer fluorographene, as obtained
  via the \textit{ab initio} polaron equations (filled triangles).
  The numbers next to some of the data points indicate the supercell dimension as a multiple of the crystalline unit
  cell. 
In (a) we also show the formation energy
  of the electron polaron, as obtained from explicit supercell calculations using SIC-DFT (circles).
In (b), the dashed line is the linear extrapolation as in Fig.~\ref{fig.extrapolation}, and the intercept with the
  vertical axis indicates the formation energy in the dilute limit. 
}
\end{figure}

\begin{figure}
  \centering
  \includegraphics[width=0.5\textwidth]{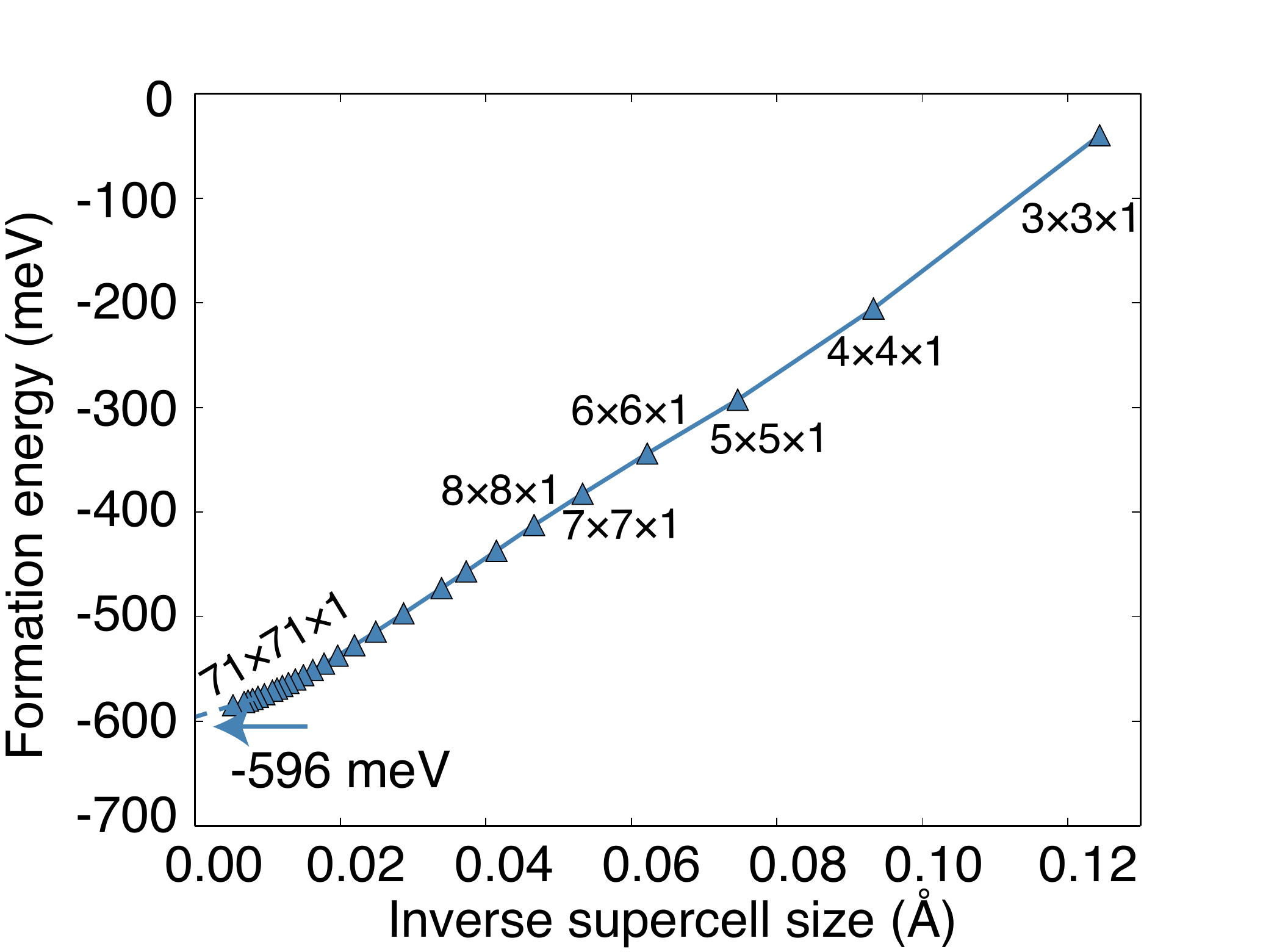}
  \caption{\label{fig.extrapol-beo}
  Formation energy of hole polaron of monolayer BeO.
  The numbers next to some of the data points indicate the supercell dimension as a multiple of the crystalline unit
  cell. The dashed line is the linear extrapolation as in Fig.~\ref{fig.extrapolation}, and the intercept with the
  vertical axis indicates the formation energy in the dilute limit.
}
\end{figure}

Figure~\ref{fig.extrapol-cf}(a) and (b) show the formation energy of electron and hole polaron in 
fluorinated graphene (CF) as a function of supercell size.
In the case of the electron polaron, the formation energy tends to zero for infinitely large supercells, indicating that 
fluorographene does not admit localized electron polarons.
In the case of holes, we find a weakly-coupled polaron, with a very small formation energy of 1.4~meV in the
dilute limit. This polaron draws weight from the $p\sigma$ states at the top of the valence band 
[Fig.~\ref{fig.cf-wfc}(a)], and is primarily driven by the $E_u$ LO modes around 40~meV [Fig.~\ref{fig.cf-wfc}(b)]. 
This very large polaron extends over more than 110$\times$110$\times$1 unit cells.

\vspace{5pt}
\textit{Calculations using self-interaction corrected DFT}

To validate our results for fluorographene obtained from the \textit{ab initio} polaron equations, we perform explicit 
supercell calculations of the electron polaron using SIC-DFT. Fig.~\ref{fig.extrapol-cf} shows that
the formation energy obtained from SIC-DFT tracks closely the results of Eqs.~(1)-(2) for large supercell
sizes, and also tends to a vanishing formation energy in the dilute limit. The deviation between the
two results at small supercell sizes, which can be seen in Fig.~\ref{fig.extrapol-cf} for 3$\times$3$\times$1 
and 4$\times$4$\times$1 supercells, is ascribed to the fact that the SIC-DFT calculations is performed using the CP 
code of Quantum ESPRESSO (modified to include SIC). Since this code uses $\Gamma$-point sampling, its calculations
converge more slowly with supercell size than Eqs.~(1)-(2). Nevertheless, the two approaches yield
practically indistinguishable results for supercells larger than 7$\times$7$\times$1 unit cells. 

\vspace{5pt}
\textit{Calculations using the 2D model}

In the case of the electron polaron, the materials parameters of CF (see Supplemental Table~S3) yield a value 
of the 2D coupling strength $\beta = 1.77$, which is below the critical strength ($\beta=2$) required for 
polaron fomation. Accordingly, Eq.~(7) of the main text predicts that no polaron will form.
This result is in agreement with the calculations using the \textit{ab initio} polaron equations and using
SIC-DFT described above.

In the case of the hole polaron, the hole mass of fluorographene 
is heavier than the electron mass (1.13~$m_e$ vs. 0.48~$m_e$).
As a result, the coupling strength for holes exceeds the critical value, $\beta = $~4.16,
and a polaron can form. Equation~(7) yields a formation energy of 0.6~meV. 
Fig.~\ref{fig.cf-wfc}(c) shows that the LO mode at 40~meV contributes 88\% of the
formation energy; therefore, when considering only the Fr\"ohlich interaction, the polaron equations
yield a formation energy of 1.2~meV, in good agreement with the 2D model.
In practice, this hole polaron is so weakly coupled that it is unlikely to be observable in experiments.

\vspace{3pt}
\textbf{3. Hole polaron in monolayer BeO}

\vspace{2pt}
\textit{Calculations using the \textit{ab initio} polaron equations}

In the case of BeO monolayer, the solution of the \textit{ab initio} polaron equations 
yields a small hole polaron. Fig~\ref{fig.beo-wfc}(a) shows that the polaron is primarily lozalized on a single
O site, with minor contributions from second-nearest-neighbor O atoms. This small polaron has a large
formation energy of 596~meV, which is a sizeable fraction of the DFT band gap of this compound (5.3~eV).
Analysis of the Fourier amplitudes of the hole wavefunction and of the atomic displacements [Fig.~\ref{fig.beo-wfc}(c)]
reveals that this particle is a small Fr\"ohlich polaron, with electronic contributions from the valence
band top at K (O-$2p$ states) and vibrational contributions
from the high-frequency LO modes and low-frequency piezoacoustic modes at long wavelength [Fig~\ref{fig.beo-wfc}(d)]. 

\vspace{5pt}
\textit{Calculations using the HSE hybrid functional}

To validate our results for monolayer BeO based on the \textit{ab initio} polaron equations, we perform explicit 
supercell calculations of the hole polaron using the HSE functional \cite{heyd2003}. This calculation is feasible 
because the small polaron fits in a 5$\times$5 supercell (50 atoms). We find that the polaron only forms when
the exchange fraction $\alpha$ in the functional exceeds the critical value $\alpha_{\rm c} =$~0.29, 
in line with our analysis in Supplemental Note~1. For $\alpha>$~0.29, 
we obtain a small hole polaron as shown in Fig.~\ref{fig.beo-wfc}(b). A comparison
between Figs.~\ref{fig.beo-wfc}(a) and (b) shows that the wavefunction produced by the \textit{ab initio} polaron 
equations, and that obtained from HSE, are practically indistinguishable. This comparison further validates
the present approach. 
For $\alpha>$~0.29, the HSE energy of the polaron increases monotonically with $\alpha$; this is
an artifact of the HSE functional as discussed in Supplemental Note 1, therefore a comparison of the energies
between HSE calculations and Eqs.~(1)-(2) is not meaningful.

\vspace{5pt}
\textit{Calculations using the 2D model}

In the case of the hole polaron in BeO, the model of Eq.~(7) yields a formation energy of 266~meV. 
In this estimate we employed the dielectric constants reported in Supplemental Table~S1.
Our explicit calculations using Eqs.~(1)-(2) indicate a small hole polaron with formation energy of 596~meV,
and the  contribution from the LO modes is of 61\%. Therefore
the formation energy resulting from the Fr\"ohlich interaction is 364~meV, which is 37\% larger than the
model hence in line with our prediction.

As for ZrS$_2$, the formation energy is sensitive to the values of the dielectric constants.
For example, by using the dielectric constants reported in Supplemental Table~S3 
instead of those in Supplemental Table~S1, we obtain a formation energy of 
218~meV, as shown in Supplemental Table~S3.

\color{black}

\clearpage
\newpage

\begin{table*}[b!]
  \begin{tabular}{l@{\hspace{0.6cm}}l@{\hspace{0.4cm}}c@{\hspace{0.4cm}}c@{\hspace{0.4cm}}c@{\hspace{0.4cm}}r}
   \toprule
    Material & $d$ (\AA) & $\epsilon_\infty$ & $\epsilon_0$ & $m_h^*$ & $m_e^*$   \\
   \midrule
    monolayer h-BN     & 2.65 &  \phantom{1}5.70 & \phantom{1}7.92 & 0.65 & 0.94  \\
    monolayer ZrS$_2$  & 5.22 &  12.25 & 29.61 & 0.29 & 0.30 \\
    fluorographene     & 4.84 &  \phantom{1}3.44 &  \phantom{1}3.87 & 1.11 & 0.48  \\
    monolayer BeO      & 3.11 &  \phantom{1}2.74 &  \phantom{1}4.38 & 3.34 & 0.83  \\
  \bottomrule
  \end{tabular}
  \caption{
    Effective thickness ($d$), dielectric constants ($\epsilon_\infty$, $\epsilon_0$),
    and effective masses ($m^*_h$, $m^*_e$) of the monolayer 
    compounds considered in this work. The dielectric constants are defined unambiguously by requiring that the screening 
    in the layer be isotropic, following Ref.~\onlinecite{Sohier2016}. To this aim, we compute the
    dielectric permittivities of each supercell, and extract the effective thickness and layer permittivities
    using Eqs.~(64) and (65) of Ref.~\onlinecite{Sio2022}.
   }
\end{table*}

\color{black}

\clearpage
\newpage

\begin{table}
  \normalsize
  \begin{tabular}{l@{\hspace{0.8cm}}l@{\hspace{0.8cm}}l}
    \toprule
    \\[-10pt]
    &
      \hspace{1.5cm}3D
    &
      \hspace{2.5cm}2D
    \\[5pt]
    \hline \\[-8pt]
     Matrix element
    &
      $\displaystyle |g(q)|^2 = E_{\rm Ha} \hbar\w a_0\,\frac{2\pi}{\Omega\k}\frac{1}{q^2}$
    &
      $\displaystyle |g(q)|^2 =  E_{\rm Ha} \hbar\w a_0\,\frac{\pi \epsilon_{\rm ion} (q_c d)^2}{2 Ad }
      \frac{1}{(q_c+q)^2}$
    \\[12pt]
    &&
       $\displaystyle q_c d = 4 \epsilon_{\infty} /(2\epsilon_{\infty}^2 \!-\!1)$ \\[8pt]
    \hline \\[-8pt]
     Effective potential
    &
      $\displaystyle V_{\rm 3D}(r) = -\frac{E_{\rm Ha} a_0}{\kappa} \frac{1}{r}$
    & 
      $\displaystyle V_{\rm 2D}(r) = - \frac{1}{2}E_{\rm Ha}\, a_0\, d q_c^2\, \epsilon_{\rm ion} \,\phi(q_c r)$
    \\[10pt]
    &&
      $\displaystyle \phi(s) = s + \frac{\pi}{2}\left[ H_0(s)-Y_0(s)\right] - \frac{\pi}{2} s\left[  H_1(s)- Y_1(s)\right]$
    \\[10pt]
    \hline \\[-7pt]
     Wavefunction ansatz
    &
      $\displaystyle \psi = (\pi r_{\rm p}^3)^{-1/2} \exp(-|\br|/r_{\rm p})$
    &
      $\displaystyle \psi = (2/\pi r_p^2)^{1/2}\exp(-|\br|/r_p)$
    \\[10pt]
    \hline \\[-8pt]
    Formation energy
    &
    $\displaystyle E = \frac{\hbar^2}{2m^* r_{\rm p}^2} - \frac{5}{16}\frac{1}{\kappa}\frac{e^2}{4\pi\varepsilon_0 r_{\rm p}}$
    &
    $\displaystyle E = \frac{\hbar^2}{2 m^* r_p^2} - \frac{\epsilon_{\rm ion}}{4}\frac{e^2 d}{4\pi \varepsilon_0  r_p^2}\, f(q_c r_p)$
   \\[13pt]
   &&
    $\displaystyle f(x) = x^2[ x^6 -\pi x^5 +28x^4 -40 \pi x^3 - 16 (1+20\log 2) x^2 $
   \\[5pt]
   &&
    $\displaystyle \quad + 240\pi x + 64(4 \log 2\! -\!7) +64 (5x^2\!-\!4)\log x ]/ (x^2+4)^4$
    \\[6pt]
    \bottomrule
\end{tabular}
  \caption{
    Comparison between the matrix elements, effective potentials, wavefunction ansatz, and formation energies
    in the Landau-Pekar model in 3D \cite{Devreese2020,Sio2019b} and the present 2D model.
    $H_0$ and $H_1$ are Struve functions, and $Y_0$ and $Y_1$ are Bessel functions of the second kind.
    In both cases, the formation energy is obtained by replacing the variational ansatz inside the 
    functional
    $
    \displaystyle E = \!\frac{\hbar^2}{2m^*}\!\!\!\int \!d\br \,|\nabla \psi|^2 +
    \!\frac{1}{2}\int \!\!d\br \,d\br'\, |\psi(\br)|^2 V(\br-\br') |\psi(\br')|^2\!,
    $
    where the integral is in three dimensions or two dimensions, respectively, 
    and $V(r)$ is replaced by $V_{\rm 3D}$ or $V_{\rm 2D}$ given in the table.
}    
\end{table}

\color{black}

\clearpage
\newpage

\begin{table*}[b!]
  \begin{tabular}{l@{\hspace{0.3cm}}l@{\hspace{0.8cm}}c@{\hspace{0.3cm}}c@{\hspace{0.4cm}}c@{\hspace{0.4cm}}c@{\hspace{0.8cm}}rrr@{\hspace{0.8cm}}r@{\hspace{0.4cm}}r}
   \toprule
   & Material & $m^*$ ($m_e$) & $\epsilon_\infty$ & $\epsilon_0$ & $d$ (\AA) & 
      $q_c$ (\AA$^{-1}$) & $\beta$ & $\gamma \cdot 10^3$ & $E$ (meV) & $r_p$ (\AA)\\
   \midrule
   Holes 
   & h-BN          & 0.65 &  \phantom{3}5.0 & \phantom{3}6.9 & 3.12 & 0.131 &  7.28 &  7.37 & -10.8 & 15.8 \\
   & MoS$_2$       & 0.57 & 15.8 & 15.9 & 6.09 & 0.021 &    0.66 &   0.21 &      -  &   -    \\
   & MoSe$_2$      & 0.65 & 17.2 & 17.7 & 6.48 & 0.018 &    3.98 &   0.14 &     0.0 &  275.4 \\
   & MoTe$_2$      & 0.72 & 20.1 & 21.9 & 7.08 & 0.014 &   17.34 &   0.08 &    -1.1 &   63.1 \\
   & WS$_2$        & 0.42 & 14.4 & 14.4 & 6.10 & 0.023 &    0.00 &   0.35 &      -  &   -    \\
   & WSe$_2$       & 0.45 & 15.6 & 15.9 & 6.47 & 0.020 &    1.65 &   0.25 &      -  &   -    \\
   & HfS$_2$       & 0.44 & 10.4 & 48.9 & 5.72 & 0.034 &  183.11 &   0.73 &  -646.0 &    5.3 \\
   & HfSe$_2$      & 0.23 & 13.9 & 71.2 & 6.11 & 0.024 &  152.17 &   0.68 &  -458.0 &    8.4 \\
   & ZrS$_2$       & 0.26 &  9.9 & 37.2 & 5.71 & 0.036 &   76.59 &   1.36 &  -313.1 &    8.6 \\
   & InSe          & 2.60 &  8.8 & 11.2 & 8.00 & 0.029 &   94.34 &   0.09 &   -28.3 &    9.3 \\
   & CF            & 1.13 & 3.13 & 3.48 &5.57 & 0.121 & 4.16 &
                     3.62 & -0.6 & 37.7 \\
   & BeO            & 3.52 & 2.74 & 4.18 & 3.03 & 0.258 & 29.02 &
                      5.30 &  -217.9 & 2.3 \\
   & AlN            & {1.49} & {4.46} & {7.31} & {2.09} & {0.220} & {16.77} &
                      {9.10} & {-119.8} & {4.2} \\
   & GaN            & {1.35} & {6.17} & {9.05} & {2.45} & {0.134} &{18.00} &
                      {3.73} &  {-57.4} & {6.4} \\
   \midrule
   Electrons &
     h-BN        & 0.83 &  \phantom{3}5.0 & \phantom{3}6.9 & 3.12 & 0.131 &     9.30 &     5.77 &   -17.5 &     12.1 \\
   & MoS$_2$    & 0.45 & 15.8 & 15.9 & 6.09 & 0.021 &    0.52 &     0.27 &    -    &   -    \\
   & MoSe$_2$   & 0.54 & 17.2 & 17.7 & 6.48 & 0.018 &    3.31 &     0.17 &     0.0 &  408.3 \\
   & MoTe$_2$   & 0.56 & 20.1 & 21.9 & 7.08 & 0.014 &   13.49 &     0.10 &    -0.8 &   78.5 \\
   & WS$_2$     & 0.31 & 14.4 & 14.4 & 6.10 & 0.023 &    0.00 &     0.47 &    -    &   -    \\
   & WSe$_2$    & 0.34 & 15.6 & 15.9 & 6.47 & 0.020 &    1.25 &     0.32 &    -    &   -    \\
   & HfS$_2$    & 0.24 & 10.4 & 48.9 & 5.72 & 0.034 &   99.88 &     1.33 &  -469.0 &    7.6 \\
   & HfSe$_2$   & 0.18 & 13.9 & 71.2 & 6.11 & 0.024 &  119.09 &     0.87 &  -402.4 &    9.8 \\
   & ZrS$_2$    & 0.31 &  9.9 & 37.2 & 5.71 & 0.036 &   91.32 &     1.14 &  -349.0 &    7.7 \\
   & InSe       & 0.10 &  8.8 & 11.2 & 8.00 & 0.029 &    3.63 &     2.29 &    -0.2 &  208.1 \\
   & {CF}            & {0.48} & 3.13 & 3.48 & 5.57 & 0.121 & 1.77 &
                      {8.53} &  - & - \\
   & {BeO}            & {0.83} & {2.74} & {4.18} & {3.03} & {0.258} & {6.84} &
                      {22.47} &  {-26.9} & {8.6} \\
   & {AlN}            & {0.51} & {4.46} & {7.31} & {2.09} & {0.220} & {5.74} &
                      {26.60} &  {-17.2} & {12.6} \\
   & {GaN}            & {0.24} & {6.17} & {9.05} & {2.45} & {0.134} & {3.20} &
                      {20.97} &  {-0.7} & {59.4} \\
  \bottomrule
  \end{tabular}
  \caption{
    Polaron formation energy $E$ and radius $r_p$ of 2D materials, as estimated using Eq.~(7) of the main text.
    The materials parameters $\beta$, $\gamma$, and $q_c$ have been obtained from the definitions given in the
    main text, using calculated effective masses and dielectric constants from the literature.
    The hole effective masses correspond to calculations for the monolayers, and are taken from the following references:
    h-BN: Ref.~\onlinecite{Sio2022}; 
    MoS$_2$, MoSe$_2$, MoTe$_2$, WS$_2$, WSe$_2$: Ref.~\onlinecite{Sohier2016};
    HfS$_2$: Ref.~\onlinecite{Wang2021};
    HfSe$_2$: Ref.~\onlinecite{Zhang2016};
    ZrS$_2$: Ref.~\onlinecite{Lv2016};
    InSe: Ref.~\onlinecite{Li2019} ($\beta$ phase);
    CF: Ref.~\onlinecite{Sivek2012};
    BeO: Ref.~\onlinecite{Ge2020};
    AlN: Ref.~\onlinecite{Cai2020};
    GaN: Ref.~\onlinecite{Jia2020}.
    The electron effective masses correspond to calculations for the monolayers, and are taken from the following references:
    h-BN: Ref.~\onlinecite{Ferreira2019};
    MoS$_2$, MoSe$_2$, MoTe$_2$, WS$_2$, WSe$_2$: Ref.~\onlinecite{Sohier2016};
    HfS$_2$, HfSe$_2$, ZrS$_2$: Ref.~\onlinecite{Zhang2014};
    InSe: Ref.~\onlinecite{Li2019}. ($\beta$ phase);
    CF: Ref.~\onlinecite{Sivek2012};
    BeO: Ref.~\onlinecite{Mortazavi2021};
    AlN: Ref.~\onlinecite{Cai2020};
    GaN: Ref.~\onlinecite{Jia2020}.
    In the case of HfS$_2$, HfSe$_2$, ZrS$_2$ the masses are anisotropic and we use the lowest mass reported.
    The dielectric constants are calculations for bulk crystals, in the in-plane direction, and are taken from
    Ref.~\onlinecite{Laturia2018} for all systems except for ZrS2, which is from Ref.~\onlinecite{Pike2018}, 
    InSe, which is from Ref.~\onlinecite{Li2020} ($\gamma$ phase), and BeO, AlN, GaN.
    For these latter compounds, the 
    bulk crystals take the wurtzite structure, therefore the bulk dielectric constants
    are not available; in these cases we calculated the bulk dielectric constants for the metastable hexagonal phases
    with AA$^\prime$ stacking.
    In all cases, the thickness $d$ is set to the 
    calculated interlayer distance, taken from Refs.~\onlinecite{Laturia2018,Li2019}.
    Dashes ``-'' indicate that localized polaron wavefunctions do not exist.
    We note that the formation energies $E$ in the penultimate column are sensitive to the precise values of
    the static and high-frequency dielectric constants. For example, by using the values in Supplemental Table~S1,
    which we calculated using the procedure of Ref.~\onlinecite{Sohier2016}, instead of the literature values 
    reported in this table, we obtain formation energies
    of 110~meV and 104~meV for the electron and hole polarons in monolayer ZrS$_2$, and a formation energy
    of 266~meV for the hole polaron in BeO (see Supplemental Note~S7).
    This sensitivity must be taken into account when performing high-throughput screenings of the polaronic properties
    of 2D materials.
   }
\end{table*}

\clearpage
\newpage

\clearpage
\newpage

\clearpage
\newpage

\clearpage
\newpage

\clearpage
\newpage

\begin{figure}
  \centering
  \includegraphics[width=0.8\textwidth]{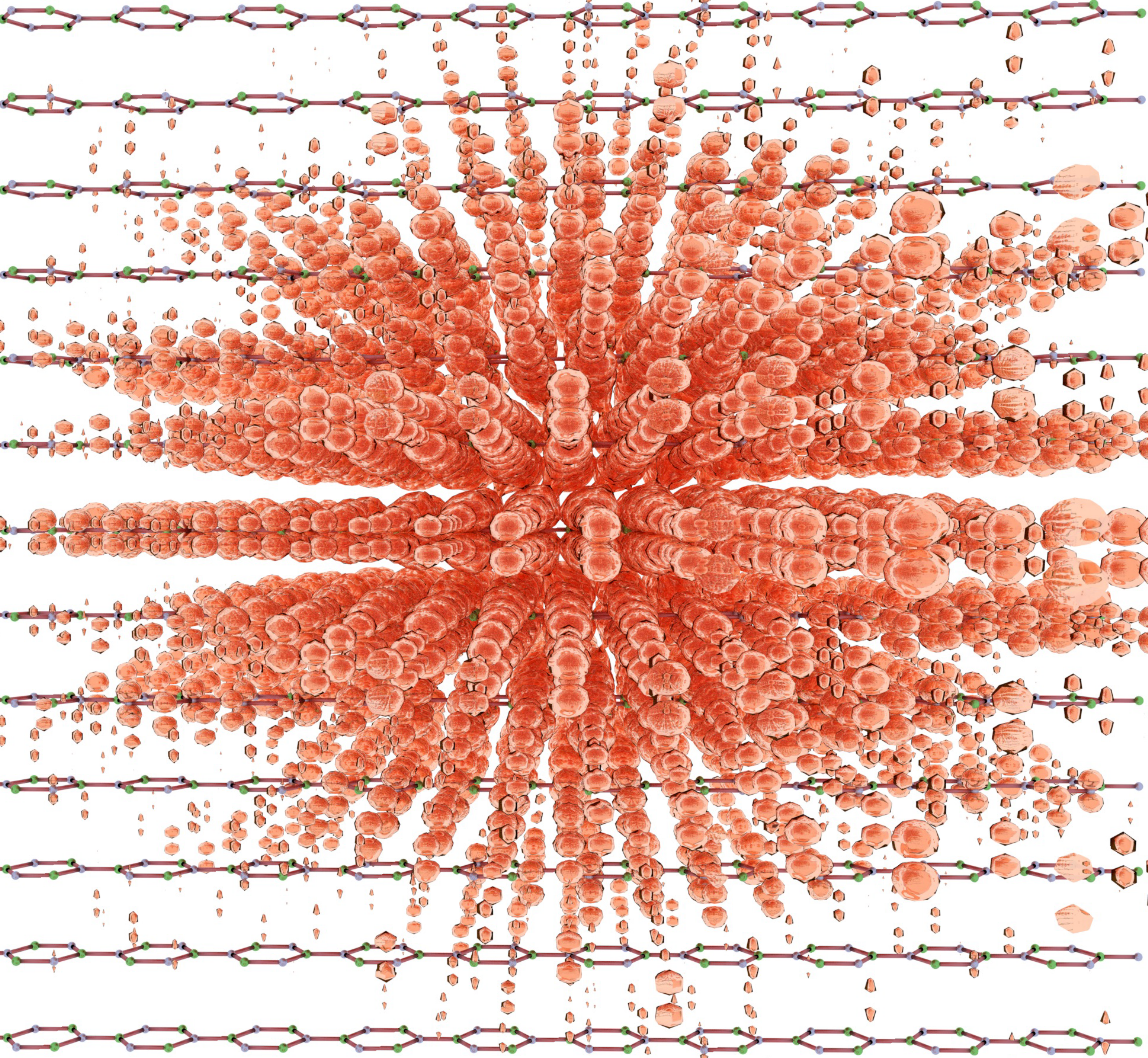}
  \caption{\label{fig.large-hbn-bulk}
  Enlarged view of the hole polaron in bulk h-BN, shown in Fig.~1(c) of the main text. The plot represents an
  isosurface of the hole polaron density $|\psi(\br)|^2$ for a 24$\times$24$\times$8 supercell (9216 atoms).
  }
\end{figure}

\clearpage
\newpage

\begin{figure}
  \centering
  \begin{tikzpicture}%[remember picture,overlay]
  \node at (0,0){\includegraphics[width=0.8\textwidth]{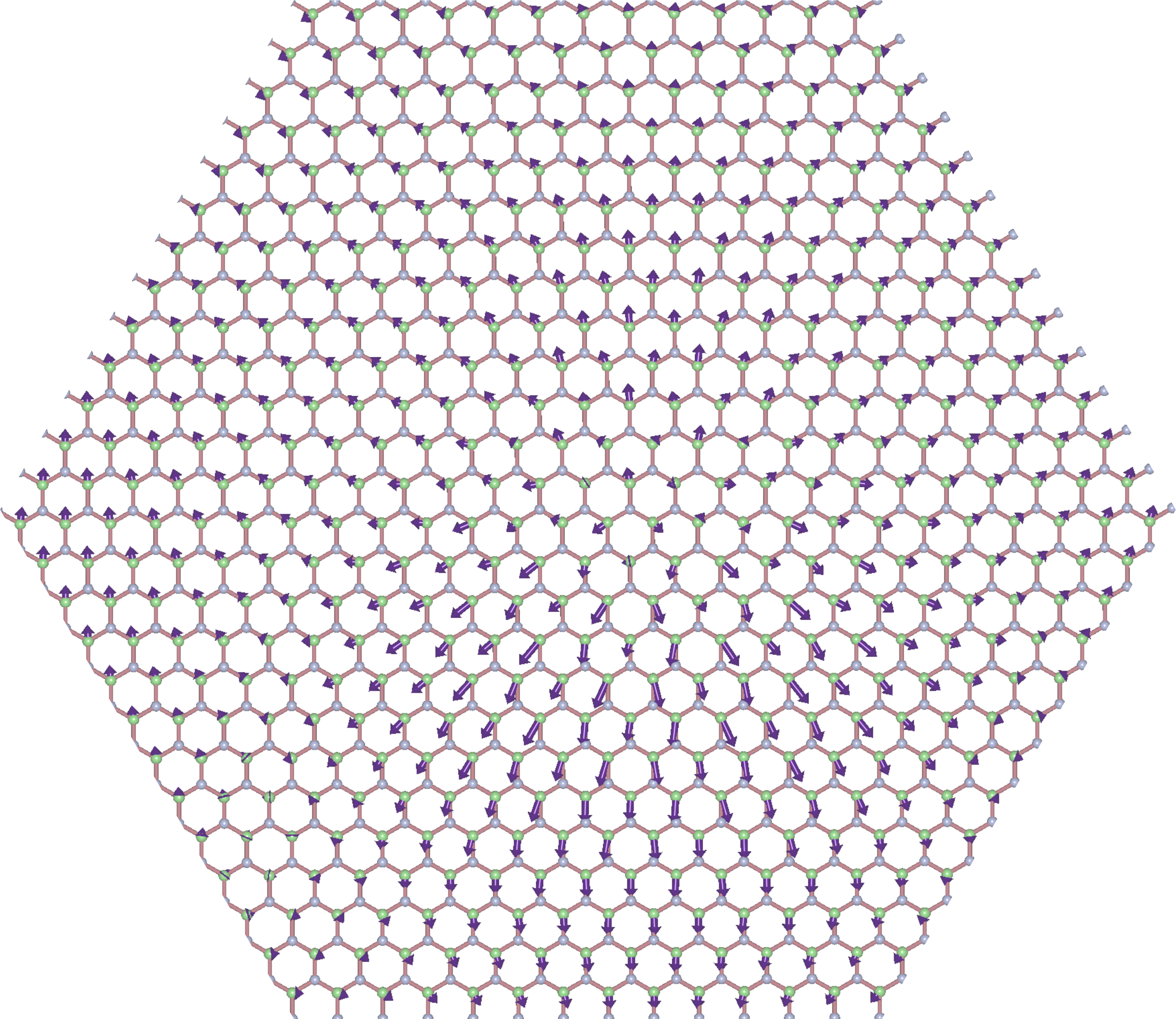}};
  \draw[fill=orange,opacity=0.4,draw=none] (0.5,0.1) circle (1cm);
  \end{tikzpicture}
  \caption{\label{fig.displ-hbn}
  Atomic displacements of the boron atoms associated with the hole polaron state in monolayer h-BN. This
  calculation corresponds to a 26$\times$26$\times$1 supercell. The arrows have been exaggerated ($\times$300) 
  for ease of visualization. The center of the polaron wavefunction shown in Fig.~2(d) of the main text
  is labeled by a disk.
  }
\end{figure}

\clearpage
\newpage

\begin{figure}
  \centering
  \includegraphics[width=0.7\textwidth]{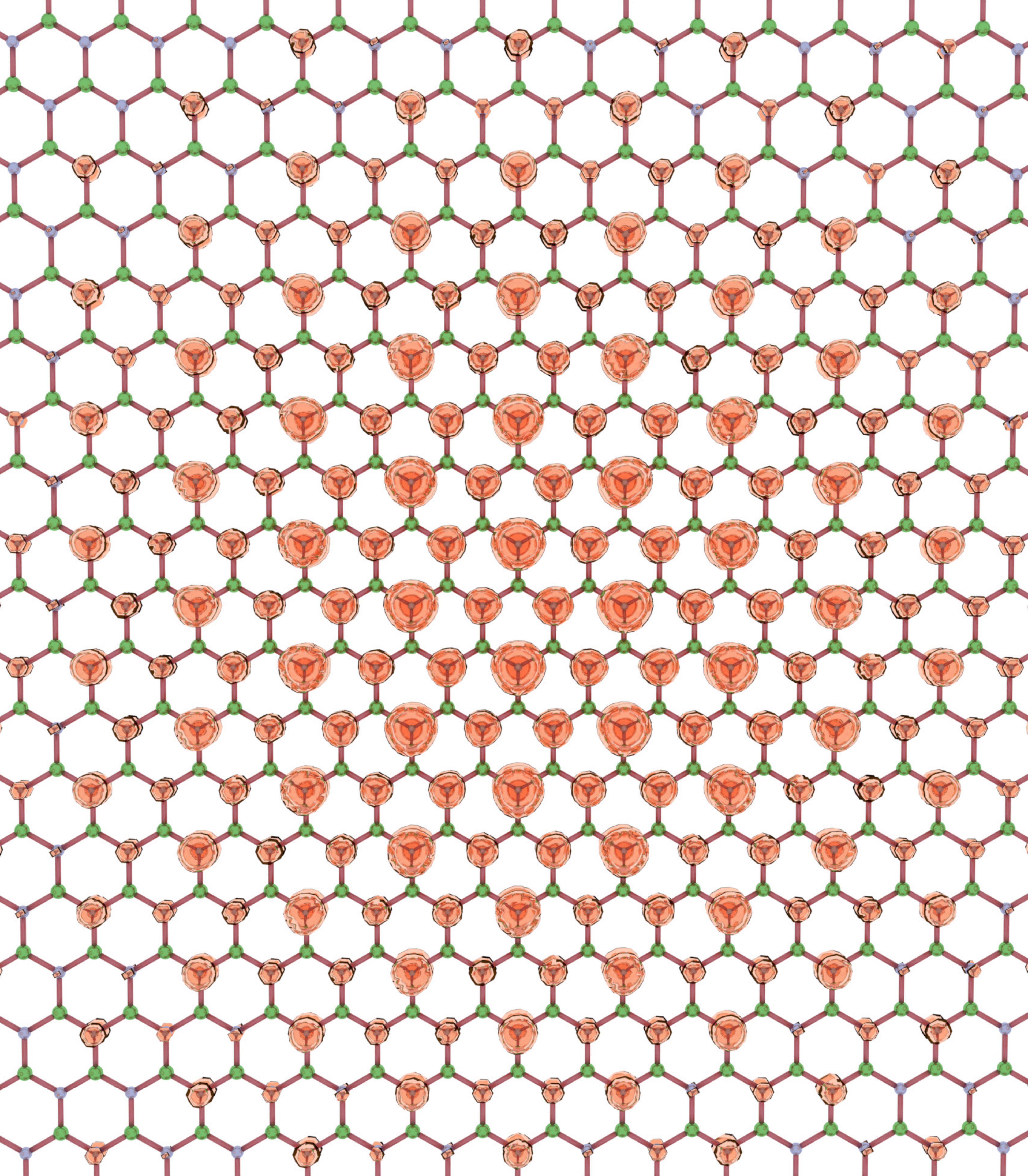}
  \caption{\label{fig.large-hbn-ml}
  Enlarged view of the hole polaron in monolayer h-BN, shown in Fig.~2(d) of the main text. The plot represents an
  isosurface of the hole polaron density $|\psi(\br)|^2$ for a 26$\times$26$\times$1 supercell (676 atoms).
  }
\end{figure}

\clearpage
\newpage

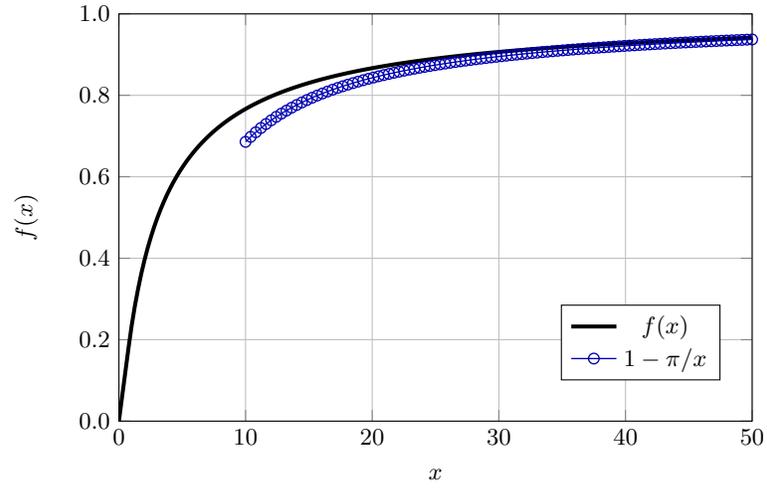
\begin{figure}
\begin{center}
    \begin{tikzpicture}
    \begin{axis}[xmin=0,xmax=50,ymin=0,ymax=1,samples=100,
               width=10cm,height=7cm,xtick={0,10,...,50},ytick={0,0.2,...,1},
      y tick label style={ /pgf/number format/.cd, fixed, fixed zerofill, precision=1, /tikz/.cd },
      x tick label style={ /pgf/number format/.cd, fixed, fixed zerofill, precision=0, /tikz/.cd },
      xlabel={$x$}, ylabel={$f(x)$},grid, legend style={ at={(0.95,0.1)}, anchor=south east}]
     \addplot+[black,line width=1.5pt,smooth,no marks,domain=0.0001:50,smooth] 
        { x^2/(x^2+4)^4*(x^6 -pi*x^5 + 28*x^4 -40*pi*x^3 - 16*(1+20*ln(2))* x^2 +240*pi*x 
         + 64*(4 *ln(2)-7) +64 *(5*x^2-4)*ln(x)) };
     \addplot+[only marks,blue!70!black,line width=0.5pt,smooth,mark=o,mark options={scale=1},domain=10:50] 
        { 1-pi/x};
     \legend{$f(x)$,$1-\pi/x$}
    \end{axis}
   % \node[] at (-1.2,4.5) {(a)};
    \end{tikzpicture}
\end{center}
\caption{\label{fig.f-function}
Plot of the dimensionless function $f(x)$ appearing in the polaron formation energy, Eq.~\eqref{eq.ef2} 
(black solid line) and its asymptotic expansion $1-\pi/x$ (blue circles).}
\end{figure}

\clearpage
\newpage

\begin{figure}
  \centering
  \includegraphics[width=\textwidth]{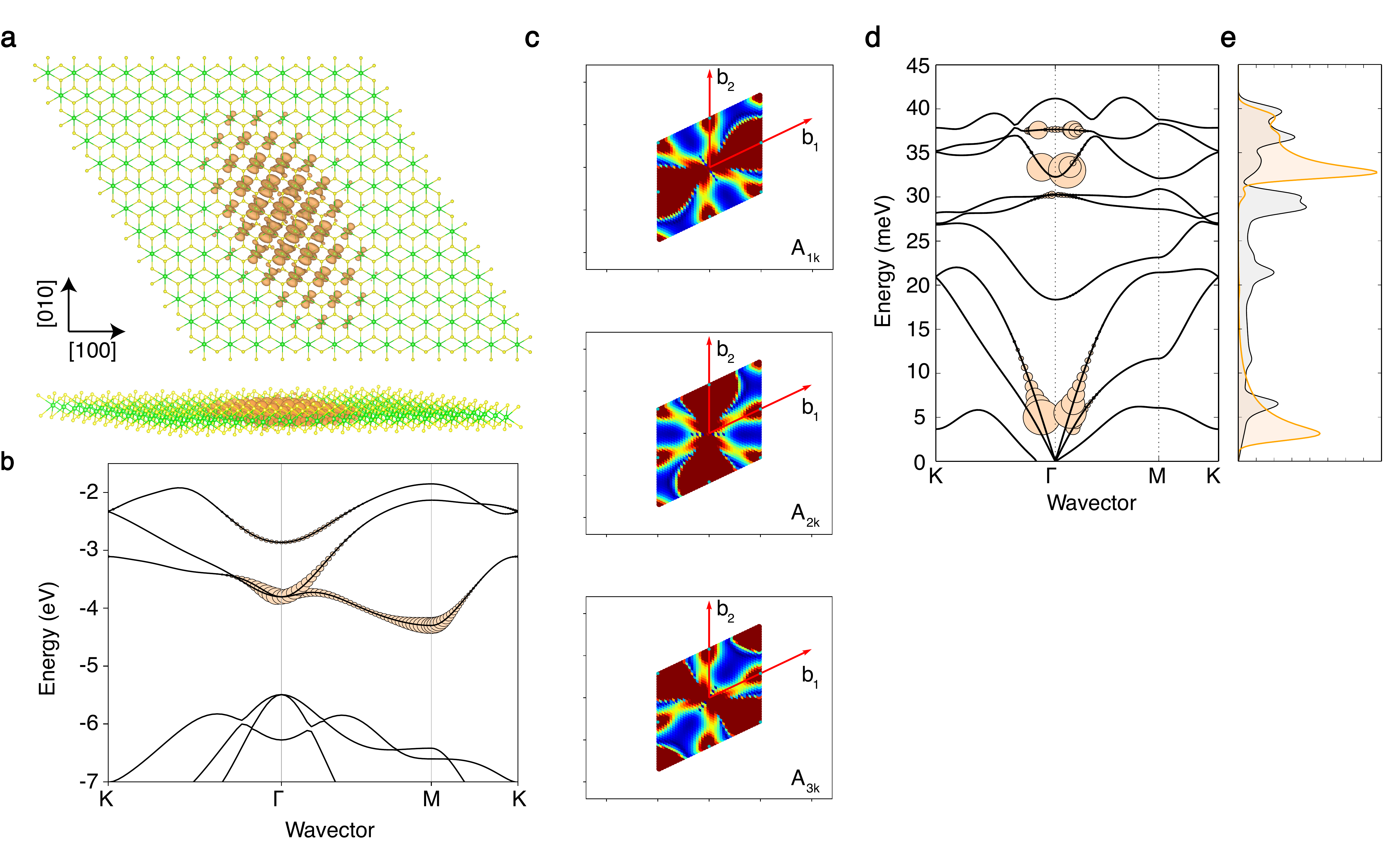}
  \caption{\label{fig.zrs2-wfc}
  Electron polaron in monolayer ZrS$_2$. (a) Isosurface plot of the polaron wavefunction, with side view and top view.
  (b) Band structure of monolayer ZrS$_2$, with the Fourier amplitudes $A_{n{\bf k}}$ of the polaron superimposed
  as circles. The radius of the circles is proportional to $|A_{n{\bf k}}|^2$. (c) 2D colormap plot of the coefficients
  $|A_{n{\bf k}}|^2$ for the three degenerate polaron states, showing threefold rotation symmetry.
  (d) Phonon dispersions with the Fourier amplitudes $B_{{\bf q}\nu}$ superimposed as circles. (e)
  Phonon density of states (black) and spectral decomposition of the vibrational contribution to the formation energy
  (orange).
}
\end{figure}

\clearpage
\newpage

\begin{figure}
  \centering
  \includegraphics[width=\textwidth]{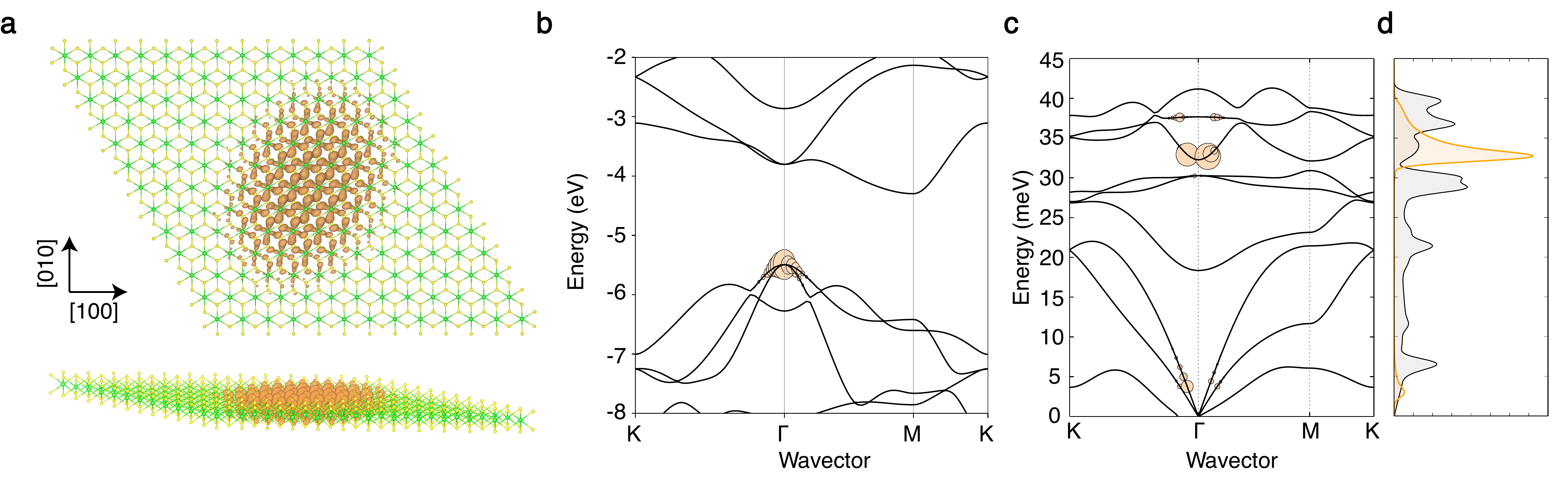}
  \caption{\label{fig.zrs2-wfc-hole}
  Hole polaron in monolayer ZrS$_2$. (a) Isosurface plot of the polaron wavefunction, with side view and top view.
  (b) Band structure of monolayer ZrS$_2$, with the Fourier amplitudes $A_{n{\bf k}}$ of the polaron superimposed
  as circles. (c) Phonon dispersions with the Fourier amplitudes $B_{{\bf q}\nu}$ superimposed as circles. (d)
  Phonon density of states (black) and spectral decomposition of the vibrational contribution to the formation energy
  (orange).
}
\end{figure}

\clearpage
\newpage

\begin{figure}
  \centering
  \includegraphics[width=\textwidth]{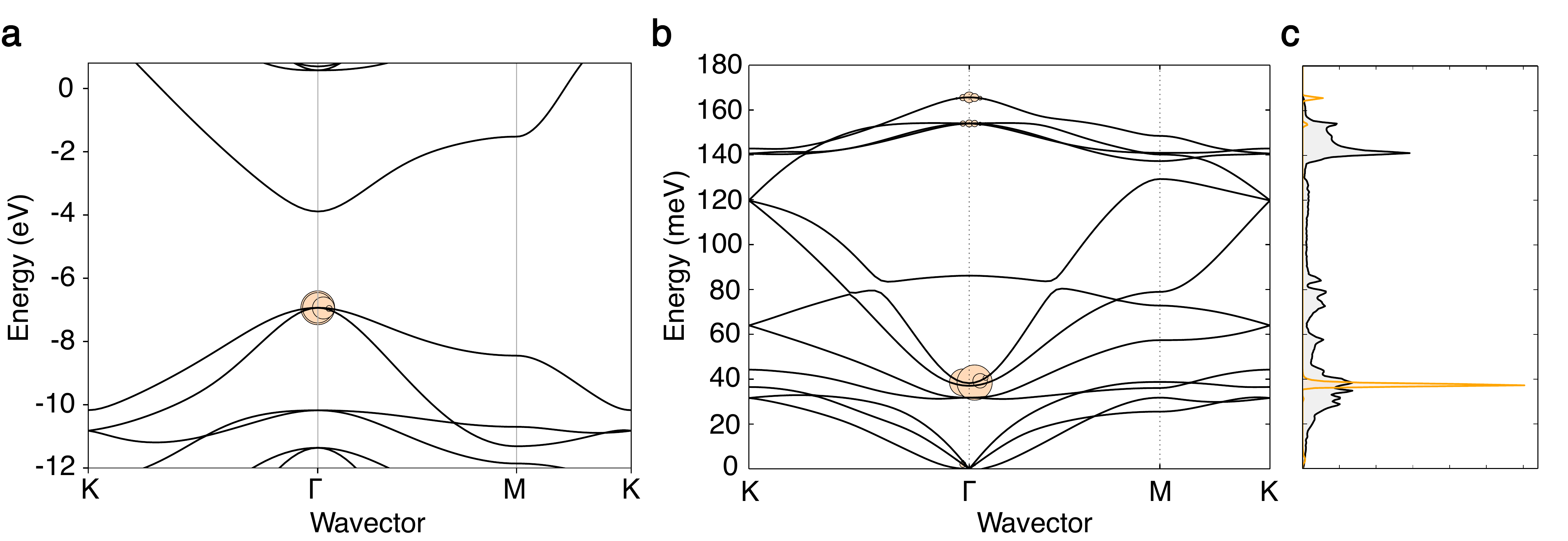}
  \caption{\label{fig.cf-wfc}
  Hole polaron in fluorographene. (a) Band structure of monolayer CF, with the Fourier amplitudes $A_{n{\bf k}}$ 
  of the polaron superimposed as circles. (b) Phonon dispersions of monolayer CF, with the Fourier amplitudes $B_{{\bf q}\nu}$ 
  superimposed as circles. (c) Phonon density of states (black) and spectral decomposition of the vibrational 
  contribution to the formation energy (orange).
}
\end{figure}

\clearpage
\newpage

\begin{figure}
  \centering
  \includegraphics[width=\textwidth]{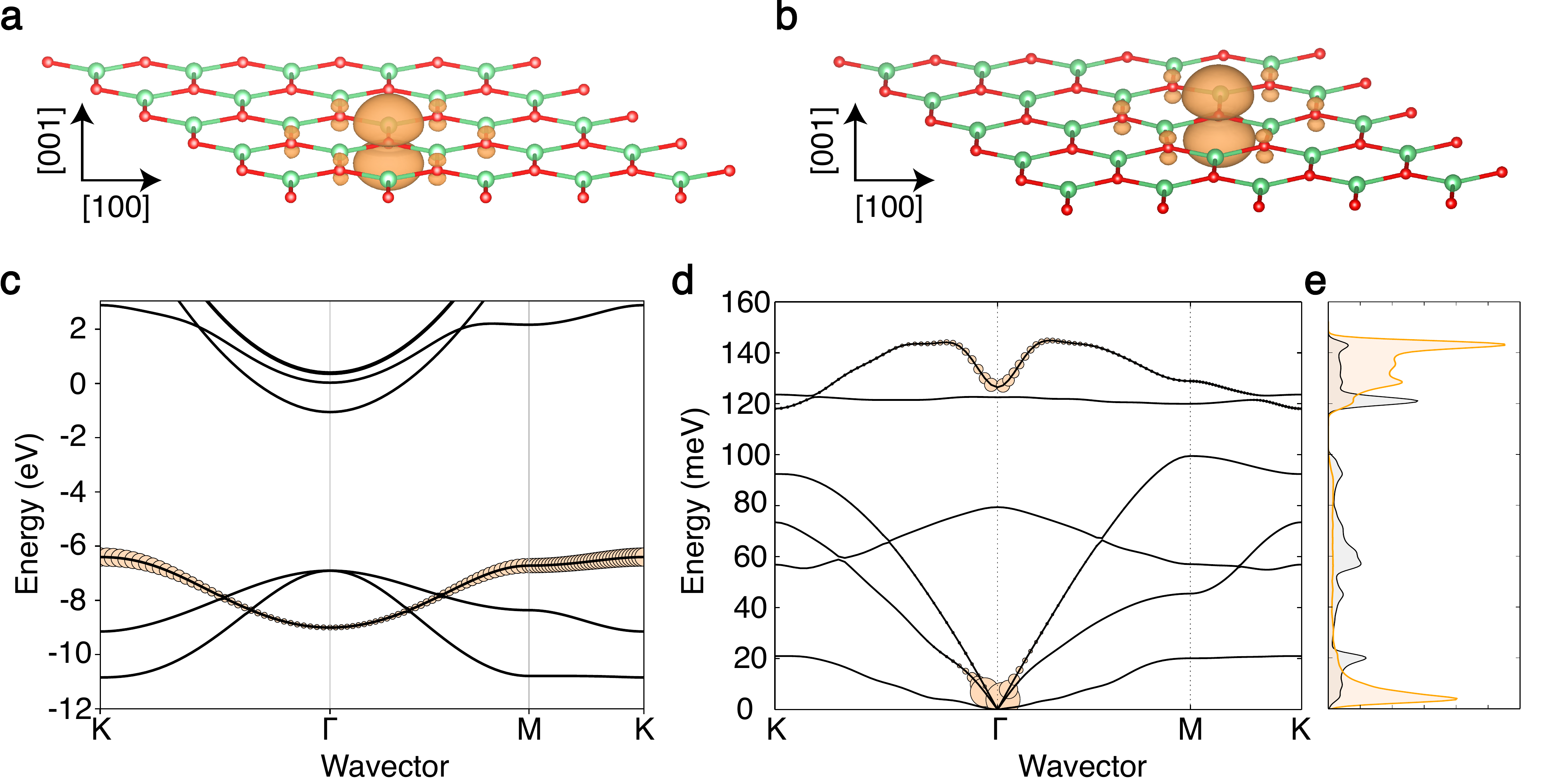}
  \caption{\label{fig.beo-wfc}
  Hole polaron in monolayer BeO. (a) Polaron wavefunction as obtained from the solution of the \textit{ab initio}
  polaron equations. (b) Polaron wavefunction as obtained by explicit supercell calculations using the HSE hybrid
  functional, with $\alpha = 0.37$. (c) Band structure of monolayer BeO, with the Fourier amplitudes $A_{n{\bf k}}$ 
  of the polaron superimposed as circles. (d) Phonon dispersions of monolayer BeO, with the Fourier amplitudes 
  $B_{{\bf q}\nu}$ superimposed as circles. (e) Phonon density of states (black) and spectral decomposition of the 
  vibrational contribution to the polaron formation energy (orange).
}
\end{figure}

\clearpage
\newpage
\bibliography{refs.bib}

\end{document}